\documentclass[12pt]{article}

\usepackage{amsmath, amsthm, amsfonts, amssymb}
\usepackage{mathrsfs} 
\usepackage[curve]{xypic}
\usepackage{graphics}

\usepackage{ifpdf}
\ifpdf
\usepackage{epstopdf,hyperref}
\else
\usepackage[hypertex]{hyperref}
\fi

\advance\textheight by \topskip
\newtheorem{Thm}{Theorem}

\newtheorem{Lem}[Thm]{Lemma}
\newtheorem{Cor}[Thm]{Corollary}
\theoremstyle{definition}
\newtheorem{Rem}[Thm]{Remark}
\newtheorem{Def}[Thm]{Definition}
\DeclareMathOperator{\cok}{cok}
\DeclareMathOperator{\End}{End}
\DeclareMathOperator{\Hom}{Hom}
\DeclareMathOperator{\hotimes}{\hat\otimes}
\DeclareMathOperator{\id}{id}
\DeclareMathOperator{\Id}{Id}

\DeclareMathOperator{\Rep}{Rep}
\DeclareMathOperator{\tr}{tr}

\numberwithin{equation}{section}
\numberwithin{Thm}{section}

\newcommand\void[1]       {}

\newcommand\arxiv[2]      {\href{http://arXiv.org/abs/#1}{#2}}
\newcommand\doi[2]        {\href{http://dx.doi.org/#1}{#2}}

\newcommand\be            {\begin{equation}}
\newcommand\bea           {\begin{equation}\begin{array}l\displaystyle}
\newcommand\bearll        {\begin{array}{ll}\displaystyle}
\newcommand\ee            {\end{equation}}
\newcommand\eear          {\end{array}}
\newcommand\enl           {\\[1em]\displaystyle}
\newcommand\etb           {&\!\! \displaystyle}

\newcommand\labl[1]       {\label{#1}\ee}

\newcommand\eev           {{}^\vee\hspace*{-1pt}}
\newcommand\eps           {\varepsilon}
\newcommand\Fs            {\mathsf{F}}
\newcommand\Ga[1]         {\Gamma\hspace*{-1pt}\big(#1\big)}
\newcommand\Gr            {\mathrm{K}_0}
\newcommand\Gs            {\mathsf{G}}
\newcommand\isorightarrow {\overset{\raisebox{-.3em}{\scriptsize $\!\sim$}}\rightarrow}
\newcommand\Lb            {\overline L}
\newcommand\mem           {\hspace*{-0.5em}}
\newcommand\one           {{\bf1}}

\newcommand\Rs            {\mathsf{R}}

\newcommand\Cb            {\mathbb{C}}

\newcommand\Zb            {\mathbb{Z}}

\newcommand\Cc            {\mathcal{C}}
\newcommand\Hc            {\mathcal{H}}
\newcommand\Ic            {\mathcal{I}}

\newcommand\Vc            {\mathcal{V}}

\begin{document}

\thispagestyle{empty}
\def\thefootnote{\fnsymbol{footnote}}
\begin{flushright}
KCL-MTH-09-03\\
\arxiv{0904.1122}{0904.1122 [hep-th]}
\end{flushright}
\vskip 5.0em
\begin{center}\LARGE
A Monoidal Category for Perturbed Defects\\
in Conformal Field Theory
\end{center}
\vskip 4em
\begin{center}\large
  Dimitrios Manolopoulos\footnote{Email: {\tt dimitrios.manolopoulos@kcl.ac.uk}}
  and
  Ingo Runkel\footnote{Email: {\tt ingo.runkel@kcl.ac.uk}}
\end{center}
\begin{center}
  Department of Mathematics, King's College London \\
  Strand, London WC2R 2LS, United Kingdom
\end{center}
\vskip 1em
\begin{center}
  April 2009
\end{center}
\vskip 4em
\begin{abstract}
Starting from an abelian rigid braided monoidal category $\Cc$ we define an abelian rigid monoidal category $\Cc_F$ which captures some aspects of perturbed conformal defects in two-dimensional conformal field theory. Namely, for $V$ a rational vertex operator algebra we consider the charge-conjugation CFT constructed from $V$ (the Cardy case). Then $\Cc = \Rep(V)$ and an object in $\Cc_F$ corresponds to a conformal defect condition together with a direction of perturbation. We assign to each object in $\Cc_F$ an operator on the space of states of the CFT, the perturbed defect operator, and show that the assignment factors through the Grothendieck ring of $\Cc_F$. This allows one to find functional relations between perturbed defect operators. Such relations are interesting because they contain information about the integrable structure of the CFT.
\end{abstract}

\setcounter{footnote}{0}
\def\thefootnote{\arabic{footnote}}

\newpage

\tableofcontents

\section{Introduction}\label{sec:I}

Conformal symmetry is a potent tool in the construction of two-dimensional conformal quantum field theories \cite{Belavin:1984vu}. Their infinite-dimensional symmetry algebra, the Virasoro algebra, is generated by the modes of two conserved currents: the holomorphic and anti-holomorphic part of the stress tensor. Besides such `chiral symmetries' obtained from conserved currents, in many examples the CFT also has an integrable symmetry, that is, infinite families of commuting conserved charges \cite{Bazhanov:1994ft}. Present approaches to CFT tend to favour either the conformal or the integrable symmetry, and it seems worthwhile to eventually combine these two symmetries into a single formalism.

In this paper we hope to take a step in this direction by continuing to develop the approach of \cite{Runkel:2007wd} which allows one to identify integrable structures of a CFT by studying the representation category of the chiral algebra. It is worth remarking that the idea to find questions about CFT that can be formulated on a purely categorical level, and that can then be investigated independent of whether there is an underlying CFT or not, has proved useful already in \cite{Fuchs:2001qc,tft1} (the interested reader could consult \cite{Kong:2009qw} for a brief overview).

\medskip

In \cite{Runkel:2007wd} families of conserved charges are constructed as perturbations of certain conformal defects. A conformal defect is a line of inhomogeneity
on the world sheet of the CFT, that is, a line where the fields can have discontinuities or singularities. By putting a circular defect line on a cylinder we obtain the defect operator, a linear operator on the space of states. If one considers a particular class of conformal defects (so-called topological defects) and perturbs such a defect by a particular type of relevant defect field, one obtains a family of defect operators which still commute with $L_0 + \Lb_0$, the sum of the zero modes of the holomorphic and anti-holomorphic component of  the stress tensor.
Sometimes these perturbed defect operators obey functional relations. An example is provided by the non-unitary Lee-Yang CFT, the Virasoro minimal model of central charge $c=-22/5$. There, one obtains a family of operators $D(\lambda)$, $\lambda \in \Cb$, on the space of states of the model, which obey, for all $\lambda,\mu\in\Cb$,
\be
  [L_0 + \Lb_0 , D(\lambda)] = 0
  ~~,~~~
  [D(\lambda),D(\mu)]=0
  ~~,~~~
  D(e^{2 \pi i/5} \lambda) \, D(e^{-2 \pi i/5} \lambda) = \id + D(\lambda) ~.
\labl{eq:intro-LY}
The last relation above is closely linked to the description of the Lee-Yang model via the massless limit of factorising scattering and the thermodynamic Bethe Ansatz, see e.g.\ the review \cite{Dorey:2007zx}. This example illustrates that the functional relations obeyed by perturbed defect operators encode at least part of the integrable structure of the model. In fact, the defect operator in \eqref{eq:intro-LY} (and more generally those for the $M_{2,2m+1}$ minimal models) can be understood as certain linear combinations of the chiral operators which were constructed in \cite{Bazhanov:1994ft} to capture the integrable structure of these models.

\medskip

In this paper we present a categorical structure that captures some aspects of perturbed defect operators, and in particular allows one to find functional relations such as the one in \eqref{eq:intro-LY}. We work in rational conformal field theory, so that the holomorphic fields of the model form a rational\footnote{
  \label{fn:rational}
  By `rational' we mean that the vertex operator algebra satisfies the conditions in \cite[Sect.\,1]{Huang2005}.}
vertex operator algebra $V$. We consider the `Cardy case' CFT constructed from $V$, namely the CFT with charge-conjugation modular invariant -- the conclusions in Section \ref{sec:C} contain a brief comment on how to extend the formalism to general rational CFTs. In the Cardy case the defects are labelled by representations of $V$. Denote $\Cc = \Rep(V)$. The category describing the properties of perturbed defects is called $\Cc_F$. It is an enlargement of $\Cc$ which depends on a choice of object $F \in \Cc$. Roughly speaking, $F$ is the representation of $V$ from which the perturbing field is taken, and the objects of $\Cc_F$ are pairs of an unperturbed defect together with a direction of perturbation.

Concretely, the objects in $\Cc_F$ are pairs $(R,f)$ where $R\in \Cc$ and $f : F \otimes R \rightarrow R$ is a morphism in $\Cc$. The morphisms in $\Cc_F$ are those morphisms in $\Cc$ which make the obvious diagram commute (see Definition \ref{def:CF-def} below). If in addition to being monoidal, the category $\Cc$ is also abelian rigid and braided (as it would be for $\Cc=\Rep(V)$ with $V$ a rational vertex operator algebra), then $\Cc_F$ is an abelian rigid monoidal category (Theorem \ref{thm:CF-rigid}). In particular, the Grothendieck ring $\Gr(\Cc_F)$ is well-defined. However, $\Cc_F$ is typically not braided. We will see in the example of the Lee-Yang model that there can be simple objects $(U,f)$ and $(V,g)$ in $\Cc_F$ such that $(U,f) \hotimes (V,g) \ncong (V,g) \hotimes (U,f)$, where $\hotimes$ denotes the tensor product in $\Cc_F$.

If $\Cc = \Rep(V)$, we can assign a perturbed defect operator $D[(R,f)]$ to an object $(R,f) \in \Cc_F$, provided certain integrals and sums converge (see Section \ref{sec:pert-def} below). Suppose that for two objects $(R,f), (S,g) \in \Cc_F$ the perturbed defect operators exist. Then the tensor product in $\Cc_F$ is compatible with composition of defect operators, $D[ (R,f) \hotimes (S,g) ] = D[(R,f)] D[(S,g)]$ (Theorem \ref{thm:D-prop}), and $D[(R,f)] = D[(S,g)]$ if $(R,f)$ and $(S,g)$ represent the same class in the Grothendieck ring $\Gr(\Cc_F)$ (Corollary \ref{cor:rep-Gr}). Thus, identities of the form $[(A,a)]\cdot[(B,b)] = [(C_1,c_1)] + \cdots + [(C_n,c_n)]$ in $\Gr(\Cc_F)$ will give rise to functional relations among the defect operators, such as the one quoted in \eqref{eq:intro-LY} (see Section \ref{sec:LY} for the Lee-Yang example).

\medskip

The category $\Cc_F$ has similarities to categorical structures that appear in the treatment of defects in other contexts.

In B-twisted $\mathcal{N}=2$ supersymmetric Landau-Ginzburg models, boundary conditions \cite{Kapustin:2002bi,Brunner:2003dc,Lazaroiu:2003zi} and defects \cite{Brunner:2007qu} can be described by so-called matrix factorisations. There, one considers a category whose objects are pairs: a $\Zb_2$-graded free module $M$ over a polynomial ring and an odd morphism $f : M \rightarrow M$, so that $f \circ f$ takes a prescribed value. The morphisms of this category have to make the same diagram commute as those of $\Cc_F$. And as in $\Cc_F$, the module $M$ can be interpreted as a defect in an unperturbed theory, and $f$ as a perturbation.  However, in the context of matrix factorisations one passes to a homotopy category, which is something we do not do for $\Cc_F$.

A more direct link comes from integrable lattice models. In one approach to these models, one uses the representation theory of a quantum affine algebra to construct families of commuting transfer matrices. The decomposition of tensor products of representations of the quantum affine algebra gives rise to functional relations among the transfer matrices \cite{Kuniba:1993cn,Rossi:2002,Korff:2003}. The category of finite-dimensional representations of a quantum affine algebra \cite{Chari:1991} shares a number of features with the category $\Cc_F$. For example, the tensor product of simple objects tends to be simple itself, except at specific points in the parameter space, where the tensor product is the middle term in a non-split exact sequence. To make the similarity a little more concrete, in Appendix \ref{sec:CF-Uq} we point out that the evaluation representations of $U_q(\widehat{sl}(2))$ can be thought of as a full subcategory of $\Cc_F$ for appropriate $\Cc$ and $F$.

\medskip

This paper is organised as follows. In Section \ref{sec:CD} we introduce the category $\Cc_F$ and study its properties. In this section we make no reference to conformal field theory or vertex operator algebras. The relation of $\Cc_F$ to defect operators in conformal field theory is described in Section \ref{sec:def}. There, we also show that the assignment of defect operators to objects in $\Cc_F$ factors through the Grothendieck ring of $\Cc_F$. In Section \ref{sec:LY} we study the Lee-Yang Virasoro minimal model conformal field theory in some detail. Section \ref{sec:C} contains our conclusions.

\bigskip\noindent
{\bf Acknowledgements:}
We would like to thank
Nils Carqueville,
J\"urgen Fuchs,
Andrew Pressley,
Chris\-toph Schweigert,
Carl Stigner,
G\'erard Watts,
and
Robert Weston
for helpful discussions and useful comments on a draft of this paper.
DM is supported by the STFC Studentship PPA/S/S/2007/04644 and
IR is partially supported by the EPSRC First Grant EP/E005047/1 and the STFC Rolling Grant ST/G000395/1.

\newpage

\section{Category theory
for perturbed defects}\label{sec:CD}

In this section we start from a monoidal category $\Cc$ and enlarge it to a new category $\Cc_F$, depending on an object $F \in \Cc$. We then investigate how properties of $\Cc$ carry over to $\Cc_F$. In particular we will see that if $\Cc$ is braided and additive then we can define a monoidal structure on $\Cc_F$. The relation to perturbed defects is discussed in more detail in Section \ref{sec:def}. The basic idea is that an object in $\Cc_F$ gives an unperturbed defect together with a direction for the perturbation by a defect field in the representation $F$.

\subsection[The category $\mathcal{C}_F$]{The category $\boldsymbol{\mathcal{C}_F}$}

\begin{Def}\label{def:CF-def}
Let $\Cc$ be a monoidal category and let $F \in \Cc$. The category $\Cc_F$ has as objects pairs $U_f \equiv (U,f)$, where $U \in \Cc$ and $f : F \otimes U \rightarrow U$. The morphisms $a : U_f \rightarrow V_g$ are all morphisms $a : U \rightarrow V$ in $\Cc$ such that the following diagram commutes:
$$
\begin{xy}
(0,15)*+{F\otimes U}="FU"; (30,15)*+{F\otimes V}="FV";(0,0)*+{U}="U"; (30,0)*+{V}="V";%
{\ar "U";"V"}?*!/_.6em/{a};%
{\ar "FU";"FV"}?*!/_.6em/{\id_F\otimes a};%
{\ar "FU";"U"}?*!/_.6em/{f};
{\ar "FV";"V"}?*!/_.6em/{g};
\end{xy}
$$
The identity morphism $\id_{U_f}$ is $\id_U$ in $\Cc$, and the composition of morphisms is that of $\Cc$.
\end{Def}

\begin{Rem}\label{rem:CF}
(i) The condition which singles out the subset of morphisms in $\Cc$ that belong to $\Cc_F$ is linear. Therefore, if $\Cc$ is an Ab-category (i.e.\ each Hom-set is an additive abelian group and composition is bilinear), then so is $\Cc_F$. Similarly, if $\Cc$ is $\Bbbk$-linear for some field $\Bbbk$, then so is $\Cc_F$.
\\[.3em]
(ii) There is an action of the monoid $\End(F)^\text{op}$ on $\Cc_F$. Namely, for each $\varphi \in  \End(F)$ we define the endofunctor $\mathcal{R}_\varphi$ of $\Cc_F$ on objects as $\mathcal{R}_\varphi(U_f) = (U , f \circ (\varphi \otimes \id_U))$ and on morphisms $U_f \overset{a}{\rightarrow} V_g$ as $\mathcal{R}_\varphi(a) = a$.
We have $\mathcal{R}_\varphi \circ \mathcal{R}_\psi = \mathcal{R}_{\psi \circ \varphi}$ without the need for natural isomorphisms.
This also shows that we have an action of $\End(F)^\text{op}$ instead of $\End(F)$.
If $\Cc$ is $\Bbbk$-linear, in this way we in particular obtain an action of $\Bbbk$ via $\lambda \mapsto \mathcal{R}_{\lambda \id_F}$.
\\[.3em]
(iii) If $\Cc$ is an Ab-category, we obtain an embedding $I$ of $\Cc$ into $\Cc_F$. The functor $I : \Cc \rightarrow \Cc_F$ is defined via $I(U) = (U,0)$ and $I(f) = f$; it is full and faithful. The forgetful functor $\Cc_F \rightarrow \Cc$ is a left inverse for $I$.
\\[.3em]
(iv) One way to think of $\Cc_F$ is as a category of `$F$-modules in $\Cc$', where the morphism $f : F \otimes U \rightarrow U$ in $U_f$ is the `action', and the morphisms of $\Cc_F$ intertwine this action. But $F$ is not required to carry any additional structure, and so there is no restriction on the `action' morphisms $f$.
\\[.3em]
(v) The category $\Cc_F$ can also be obtained as a (non-full) subcategory of the comma category $(F \otimes(-) \downarrow \Id)$ (see \cite[Sect.\,II.6]{MacLane-book} for more on comma categories). The objects of $(F \otimes(-) \downarrow \Id)$ are triples $(U,V,f)$ where $U,V \in \Cc$ and $f:F\otimes U \rightarrow V$. The morphisms $(U,V,f) \rightarrow (U',V',f')$ are pairs $(x : U \rightarrow U', y: V \rightarrow V')$ so that $y \circ f = f' \circ (\id_F \otimes x)$. The subcategory in question consists of all objects of the form $(U,U,f)$ and all morphisms of the form $(x,x)$.
\\[.3em]
(vi) The category of evaluation representations of the quantum affine algebra $U_q(\widehat{sl}(2))$ is a full subcategory of $\Rep(U_q(sl(2)))_F$, where $F$ is $U_q(\widehat{sl}(2))$ understood as a $U_q(sl(2))$-module. The details can be found in Appendix \ref{sec:CF-Uq}. As briefly mentioned in the introduction, 
short exact sequences of representations of $U_q(\widehat{sl}(2))$ provide identities between transfer matrices for certain integrable lattice models. On the other hand, in Section \ref{sec:def} below we will see that short exact sequences in $\Cc_F$ give identities between certain defect operators in CFT. We hope that this similarity can be made more concrete in the future.
\end{Rem}

Recall that the Grothendieck group $\Gr(\Cc)$ of an abelian category $\Cc$ is the free abelian group generated by isomorphism classes $(U)$ of objects $U$ in $\Cc$, quotiented by the subgroup generated by the relations $(U) = (K) + (C)$ for all short exact sequences $0 \rightarrow K \rightarrow U \rightarrow C \rightarrow 0$. We denote the equivalence class of $(U)$ in $\Gr(\Cc)$ by $[U]$. The following theorem provides a sufficient condition for $\Cc_F$ to be abelian, so that it makes sense to talk about the Grothendieck group of $\Cc_F$.
The proof is given in Appendix \ref{app:proof-ab}.

\begin{Thm} \label{thm:ab}
If $\Cc$ is an abelian monoidal category with right-exact tensor product, then $\Cc_F$ is abelian.
\end{Thm}

Recall that a functor $G:\Cc \rightarrow \mathcal{D}$ is said to be {\em right-exact} if for $U,V,W \in \Cc$, exactness of $U \rightarrow V \rightarrow W \rightarrow 0$ implies exactness of $G(U) \rightarrow G(V) \rightarrow G(W) \rightarrow 0$. A tensor product bifunctor is called right-exact if $X \otimes(-)$ and $(-)\otimes X$ are right-exact functors for all $X\in\Cc$. The following lemma will be useful; it is also proved in Appendix \ref{app:proof-ab}.

\begin{Lem} \label{lem:CF-exact}
Let $\Cc$ be as in Theorem \ref{thm:ab} and $U_f \xrightarrow{a} V_g \xrightarrow{b} W_h$ be a complex in $\Cc_F$. Then $U_f \xrightarrow{a} V_g \xrightarrow{b} W_h$ is exact at $V_g$ in $\Cc_F$ iff
$U \xrightarrow{a} V \xrightarrow{b} W$ is exact at $V$ in $\Cc$.
\end{Lem}

\subsection[Monoidal structure on $\Cc_F$]{Monoidal structure on $\boldsymbol{\Cc_F}$}\label{sec:CF-mon}

Let $\Cc$ be a braided monoidal Ab-category. Following the conventions of \cite[Sect.\,VII.1]{MacLane-book} we denote the associator by $\alpha_{U,V,W} : U \otimes (V \otimes W) \isorightarrow (U\otimes V)\otimes W$ and the unit isomorphisms by $\lambda_U : \one \otimes U \isorightarrow U$ and $\rho_U : U \otimes \one \isorightarrow U$. The braiding isomorphisms are $c_{U,V} : U \otimes V \isorightarrow V \otimes U$.

The braiding and the abelian group structure on $\Hom$-spaces allows us to define a tensor product $\hotimes$ on $\Cc_F$ as follows. On objects $U_f, V_g \in \Cc_F$ we set
\be
  U_f \hotimes V_g = (U \otimes V , T(f,g)) ~,
\ee
where $T(f,g) : F \otimes (U \otimes V) \rightarrow U \otimes V$ is defined as
\be
  T(f,g) = (f \otimes \id_V) \circ \alpha_{F,U,V} +
  (\id_U \otimes g) \circ \alpha^{-1}_{U,F,V} \circ (c_{F,U} \otimes \id_V) \circ \alpha_{F,U,V} ~.
\labl{eq:T-def}
This definition, and some of the definitions and arguments below, are easier to understand upon replacing $\Cc$ by an equivalent strict category (which has trivial associators and unit isomorphisms) and using the graphical representation of morphisms in braided monoidal categories, cf.\ \cite[Sect.\,2.3]{BaKi-book}. We use the conventions in \cite[Sect.\,2]{tft1}. For example, the graphical representation of \eqref{eq:T-def} is
\be
  T(f,g) = ~
  \raisebox{-37pt}{
  \begin{picture}(50,80)
   \put(0,5){\scalebox{.90}{\includegraphics{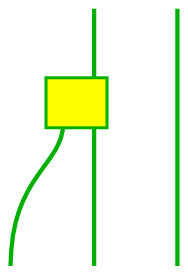}}}
   \put(0,5){
     \setlength{\unitlength}{.90pt}\put(-16,-1){
     \put( 13, -7) {\scriptsize$ F $}
     \put( 33, 47) {\scriptsize$ f $}
     \put( 37, -7) {\scriptsize$ U $}
     \put( 62, -7) {\scriptsize$ V $}
     \put( 37, 80) {\scriptsize$ U $}
     \put( 62, 80) {\scriptsize$ V $}
      }\setlength{\unitlength}{1pt}}
  \end{picture}}
  ~~+
  \raisebox{-37pt}{
  \begin{picture}(50,80)
  \put(0,5){\scalebox{.90}{\includegraphics{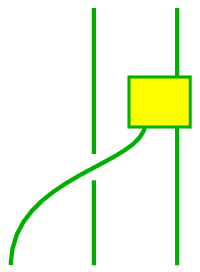}}}
  \put(0,5){
    \setlength{\unitlength}{.90pt}\put(-32,-1){
    \put( 28, -7) {\scriptsize$ F $}
    \put( 74, 48) {\scriptsize$ g $}
    \put( 53, -7) {\scriptsize$ U $}
    \put( 77, -7) {\scriptsize$ V $}
    \put( 53, 80) {\scriptsize$ U $}
    \put( 77, 80) {\scriptsize$ V $}
      }\setlength{\unitlength}{1pt}}
  \end{picture}}
~~.
\ee
We will write $\one$ for the object $\one_0 \equiv (\one,0)$ in $\Cc_F$. This will be the tensor unit for $\hotimes$.

\begin{Lem}\label{lem:CF-ass-unit}
The associator and unit isomorphisms of $\Cc$ are isomorphisms in $\Cc_F$ as follows: $\alpha_{U,V,W} : U_f \hotimes (V_g \hotimes W_h) \rightarrow (U_f \hotimes V_g) \hotimes W_h$, $\lambda_{U_f} : \one \hotimes U_f \rightarrow U_f$ and $\rho_{U_f} : U_f \hotimes \one \rightarrow U_f$.
\end{Lem}

\begin{proof}
We have to show that
\bea
  \alpha_{U,V,W} : \big(U \otimes (V \otimes W), T(f,T(g,h))\big) \rightarrow \big((U \otimes V) \otimes W,T(T(f,g),h)\big)~,
  \enl
  \lambda_U : \big(\one \otimes U,T(0,f)\big) \rightarrow (U,f) ~~,~~~
  \rho_U : \big(U \otimes \one, T(f,0)\big) \rightarrow (U,f)
\eear\ee
make the diagram in Definition \ref{def:CF-def} commute. These are all straightforward calculations. For example,
$\rho_U \circ T(f,0)
  = \rho_U \circ (f \otimes \id_\one) \circ \alpha_{F,U,\one}
  = f \circ \rho_{F \otimes U} \circ \alpha_{F,U,\one}
  = f \circ (\id_F \otimes \rho_U)$.
\end{proof}

\begin{Lem}
Let $a : U_f \rightarrow U'_{f'}$ and $b : V_g \rightarrow V'_{g'}$ be morphisms in $\Cc_F$. Then $a \otimes b : U \otimes V \rightarrow U' \otimes V'$
is also a morphism $U_f \hotimes V_g \rightarrow U'_{f'} \hotimes V'_{g'}$ in $\Cc_F$.
\end{Lem}

\begin{proof}
We have to show that $(a \otimes b)\circ T(f,g)=T(f',g')\circ (\id_F\otimes(a \otimes b))$. By \eqref{eq:T-def}, $T(f,g)$ splits into two summands as $T(f,g)=T(f,0)+T(0,g)$ (same also holds for $T(f',g')$). We start by showing that $(a \otimes b)\circ T(f,0)=T(f',0) \circ (\id_F\otimes(a \otimes b))$:
\bea
(a \otimes b)\circ T(f,0) =((a \circ f) \otimes b) \circ \alpha_{F,U,V}
\stackrel{(1)}{=}(f' \otimes \id_{V'}) \circ ((\id_F \otimes a) \otimes b)\circ \alpha_{F,U,V} \\[.4em]\displaystyle
\stackrel{(2)}{=}(f' \otimes \id_{V'})\circ \alpha_{F,U',V'} \circ (\id_F \otimes (a \otimes b)) = T(f',0) \circ (\id_F\otimes(a \otimes b)) ~.
\eear\ee
In step 
(1) we used the fact that $a \circ f=f' \circ (\id_F \otimes a)$, since $a$ is a morphism in $\Cc_F$, and step 
(2) amounts to naturality of $\alpha_{F,U,V}$ in $U$ and $V$. The proof of $(a \otimes b)\circ T(0,g)=T(0,g') \circ (\id_F\otimes(a \otimes b))$ goes along the same lines, using also that $c_{F,U}$ is natural in $U$.
\end{proof}
\medskip

According to the previous lemma, on morphisms $a,b$ we can define the tensor product to be the same as in $\Cc$,
\be
  a \hotimes b = a \otimes b ~.
\ee
One checks that $\hotimes$ is a bifunctor. Together with Lemma \ref{lem:CF-ass-unit} this shows that $\Cc_F$ is a monoidal category.

\begin{Rem}
(i) Even though $\Cc$ is braided, $\Cc_F$ is in general not. The reason is that $c_{U,V}$ is typically not a morphism in $\Cc_F$. Also, we actually demand too much when we require $\Cc$ to be braided, since all we use are the braiding isomorphisms where one of the arguments is given by $F$.
\\[.3em]
(ii) The functors $\mathcal{R}_\varphi$ defined in Remark \ref{rem:CF} are strict monoidal functors. That is, $\mathcal{R}_\varphi(U_f \hotimes V_g)$ $=$ $\mathcal{R}_\varphi(U_f) \hotimes \mathcal{R}_\varphi(V_g)$ for objects and $\mathcal{R}_\varphi(a \hotimes b) = \mathcal{R}_\varphi(a) \hotimes \mathcal{R}_\varphi(b)$ for morphisms. This follows from $T(f \circ (\varphi \otimes \id_U), g \circ (\varphi \otimes \id_V)) = T(f,g) \circ (\varphi \otimes \id_{U\otimes V})$.
\end{Rem}

\begin{Thm}\label{thm:CF-tens}
If $\Cc$ is an abelian braided monoidal category with right-exact tensor product, then $\Cc_F$ is an abelian monoidal category with right-exact tensor product. If the tensor product of $\Cc$ is exact, then so is that of $\Cc_F$.
\end{Thm}

\begin{proof}
We have seen above that $\Cc_F$ is monoidal and in Theorem \ref{thm:ab} that $\Cc_F$ is abelian. We will show that if $\otimes$ is right-exact, then the functor $X_x \hotimes(-)$ is right-exact. The arguments for $(-) \hotimes X_x$ and for `exact' in place of `right-exact' are analogous. Let $U_f \xrightarrow{a} V_g \xrightarrow{b} W_h \rightarrow 0$ be exact. Then $X \otimes U \xrightarrow{\id_X \otimes a} X \otimes V \xrightarrow{\id_X \otimes b} X \otimes W \rightarrow 0$ is exact in $\Cc$. By Lemma \ref{lem:CF-exact}, $X_x \hotimes U_f \xrightarrow{\id_X \otimes a} X_x \hotimes V_g \xrightarrow{\id_X \otimes b} X_x \hotimes W_h \rightarrow 0$ is exact in $\Cc_F$.
\end{proof}

If $\Cc$ is monoidal with exact tensor product, then the Grothendieck group $\Gr(\Cc)$ carries a ring structure defined via $[U] \cdot [V] = [U \otimes V]$. In this case, $\Gr(\Cc)$ is called the {\em Grothendieck ring}.

\begin{Cor}\label{cor:Gr-exists}
If $\Cc$ is an abelian braided monoidal category with exact tensor product, then $\Cc_F$ has a well-defined Grothendieck ring $\Gr(\Cc_F)$.
\end{Cor}

\subsection[Duality on $\Cc_F$]{Duality on $\boldsymbol{\Cc_F}$}

Let $\Cc$ be a monoidal category. We say that $\Cc$ has right-duals if for each object $U$ there is an object $U^\vee$ together with morphisms $b_U : \one \rightarrow U \otimes U^\vee$, $d_U : U^\vee \otimes U \rightarrow \one$ such that
\be\begin{array}{rl}\displaystyle
  \rho_U \circ (\id_U \otimes d_U) \circ \alpha^{-1}_{U,U^\vee,U} \circ (b_U \otimes \id_U) \circ \lambda_U^{-1}
  \etb= \id_U~, \enl
  \lambda_{U^\vee} \circ (d_U \otimes \id_{U^\vee}) \circ \alpha_{U^\vee,U,U^\vee} \circ (\id_{U^\vee} \otimes b_U) \circ \rho^{-1}_{U^\vee} \etb = \id_{U^\vee}~,
\eear\labl{eq:r-dual}
see e.g.\ \cite[Sect.\,2.1]{BaKi-book}. A graphical representation of these identities can be found in \cite[Sect.\,2.3]{BaKi-book} or \cite[Eqn.\,(2.10)]{tft1}.
We say that $\Cc$ has left duals if for each object $U$ there is an object $\eev U$ together with morphisms $\tilde b_U : \one \rightarrow \eev U \otimes U$, $\tilde d_U : U \otimes \eev U \rightarrow \one$ such that
\be\begin{array}{rl}\displaystyle
  \lambda_U \circ (\tilde d_U \otimes \id_U) \circ \alpha_{U,\eev U,U} \circ (\id_U \otimes \tilde b_U) \circ \rho_U^{-1} \etb= \id_U~, \enl
  \rho_{\eev U} \circ (\id_{\eev U} \otimes \tilde d_U) \circ \alpha_{\eev U,U,\eev U}^{-1} \circ  (\tilde b_U \otimes \id_{\eev U}) \circ \lambda^{-1}_{\eev U} \etb= \id_{\eev U}
  ~.
\eear\labl{eq:l-dual}
Suppose now that $\Cc$ is a braided monoidal Ab-category which has right duals. To a given object $U_f \in \Cc_F$ we assign the object
\be \begin{array}{ll}\displaystyle
  (U_f)^\vee = (U^\vee, c(f)) ~;~~
  c(f) \etb=
  - \lambda_{U^\vee} \circ (d_U \otimes \id_{U^\vee}) \circ ( (\id_{U^\vee} \otimes f) \otimes \id_{U^\vee})
  \circ ( \alpha^{-1}_{U^\vee,F,U} \otimes \id_{U^\vee})
  \\ \etb
  \qquad \circ \alpha_{U^\vee \otimes F, U, U^\vee}
  \circ (c_{F,U^\vee} \otimes b_U) \circ (\rho_{F \otimes U^\vee})^{-1}~.
\eear\labl{eq:cf-def}
If $\Cc$ has left duals, we define analogously
\be \begin{array}{ll}\displaystyle
  \eev(U_f) = (\eev U, \tilde c(f)) ~;~~
  \tilde c(f) \etb=
  - \rho_{\eev U} \circ (\id_{\eev U} \otimes \tilde d_U) \circ \alpha^{-1}_{\eev U,U,\eev U} \circ
  ( (\id_{\eev U} \otimes (f \circ c_{F,U}^{-1})) \otimes \id_{\eev U})
  \\ \etb
  \qquad \circ
  ( \alpha^{-1}_{\eev U,U,F} \otimes \id_{\eev U} ) \circ ( (\tilde b_U \otimes \id_F) \otimes \id_{\eev U})
  \circ (\lambda_F^{-1} \otimes \id_{\eev U}) ~.
\eear\labl{eq:tilde-cf-def}
As for \eqref{eq:T-def} it is helpful to pass to a strict category and write out the graphical representation of \eqref{eq:cf-def} and \eqref{eq:tilde-cf-def}. This leads to the simple expressions
\be
  c(f) = -
  \raisebox{-37pt}{
  \begin{picture}(50,80)
  \put(0,5){\scalebox{.90}{\includegraphics{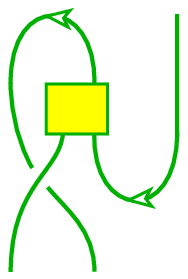}}}
  \put(0,5){
    \setlength{\unitlength}{.90pt}\put(-32,-1){
    \put( 28, -7) {\scriptsize$ F $}
    \put( 49, 47) {\scriptsize$ f $}
    \put( 53, -7) {\scriptsize$ U^{\vee} $}
    \put( 77, 80) {\scriptsize$ U^{\vee} $}
      }\setlength{\unitlength}{1pt}}
  \end{picture}}
  \quad , \quad
  \tilde c(f) = -
  \raisebox{-37pt}{
  \begin{picture}(50,80)
  \put(0,5){\scalebox{.90}{\includegraphics{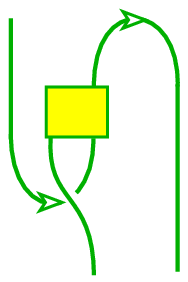}}}
  \put(0,5){
    \setlength{\unitlength}{.90pt}\put(-32,-1){
    \put( 53, -7) {\scriptsize$ F $}
    \put( 49, 47) {\scriptsize$ f $}
    \put( 75, -7) {\scriptsize$ \eev U $}
    \put( 26, 80) {\scriptsize$ \eev U $}
      }\setlength{\unitlength}{1pt}}
  \end{picture}}
  ~.
\ee

\begin{Lem}\label{lem:CF-duals}
(i) If $\Cc$ has right duals, then $b_U : \one \rightarrow U_f \hotimes (U_f)^\vee$ and
$d_U : (U_f)^\vee \hotimes U_f \rightarrow \one$ are morphisms in $\Cc_F$.
(ii) If $\Cc$ has left duals, then $\tilde b_U : \one \rightarrow \eev(U_f) \hotimes U_f$ and
$\tilde d_U : U_f \hotimes \eev(U_f) \rightarrow \one$ are morphisms in $\Cc_F$.
\end{Lem}

\begin{proof}
The proof works similar in all four cases. Consider $b_U$ as an example. The commuting diagram in Definition \ref{def:CF-def} boils down to the condition that the morphism $T(f,c(f)) \circ (\id_F \otimes b_U) : F \otimes \one \rightarrow U \otimes U^\vee$ has to be zero, i.e.\ that
\be
  T(f,0) \circ (\id_F \otimes b_U) = - T(0,c(f)) \circ (\id_F \otimes b_U) ~.
\ee
Let us suppose that $\Cc$ is strict. The non-strict case then follows by invoking coherence and verifying that the $\alpha$, $\rho$ and $\lambda$ sit in the required places. The calculation is really best done using the graphical notation, but let us write out the individual steps in equations. The right hand side of the above equation then reads
\bea
  - T(0,c(f)) \circ (\id_F \otimes b_U)
  \\[.5em]\displaystyle
  =   (\id_U \otimes d_U \otimes \id_{U^\vee}) \circ (\id_{U \otimes U^\vee} \otimes f \otimes \id_{U^\vee})
  \circ (\id_U \otimes c_{F,U^\vee} \otimes b_U)
 \circ (c_{F,U} \otimes \id_{U^\vee}) \circ   (\id_F \otimes b_U)
  \\[.3em]\displaystyle
  \overset{(1)}{=}   (\id_U \otimes d_U \otimes \id_{U^\vee}) \circ (\id_{U \otimes U^\vee} \otimes f \otimes \id_{U^\vee})
  \circ (c_{F,U \otimes U^\vee} \otimes b_U) \circ   (\id_F \otimes b_U)
  \\[.3em]\displaystyle
  \overset{(2)}{=}   (\id_U \otimes d_U \otimes \id_{U^\vee}) \circ (b_U \otimes \id_{U^\vee} \otimes \id_{U^\vee}) \circ (f \otimes \id_{U^\vee})
  \circ (\id_F \otimes b_U)
  \\[.3em]\displaystyle
  \overset{(3)}{=} (f \otimes \id_{U^\vee}) \circ (\id_F \otimes b_U) = T(f,0) \circ (\id_F \otimes b_U) ~.
\eear\ee
Step (1) amounts to one of the hexagon identities for the braiding, (2) uses naturality of the braiding to pull $b_U$ through $c_{F,U \otimes U^\vee}$ and the fact that $c_{F,\one} = \id_F$.  Step (3) is property \eqref{eq:r-dual} of the right duality.
\end{proof}

\medskip

A monoidal category is called {\em rigid} if every object has left and right duals \cite[Def.\,2.1.2]{BaKi-book}.
The above lemma immediately implies the following theorem.

\begin{Thm}\label{thm:CF-rigid}
Let $\Cc$ be a braided monoidal Ab-category. If $\Cc$ has right and/or left duals, then so has $\Cc_F$.
In particular, if $\Cc$ is rigid, so is $\Cc_F$.
\end{Thm}

\begin{Rem}
(i) Suppose $\Cc$ has left and right duals. Even if in $\Cc$ we would have $U^\vee = \eev U$, the same need not be true in $\Cc_F$ due to the distinct definitions of $c(f)$ and $\tilde c(f)$. Also, even if in $\Cc$ we would have $(U^\vee)^\vee \cong U$, the same need not hold in $\Cc_F$. We will see this explicitly in the Lee-Yang example in Section \ref{sec:LY-ex-seq}.
\\[.3em]
(ii) Let $\Cc$ be as in Corollary \ref{cor:Gr-exists}. If $\Cc$ has right duals, then the existence of a right duality for $\Cc_F$ tells us that in $\Gr(\Cc_F)$ we have $[(U_f)^\vee] \cdot [U_f] = [\one] + [W_h]$ and $[U_f] \cdot [(U_f)^\vee] = [\one] + [W'_{h'}]$ for some $W_h,W'_{h'} \in \Cc_F$. This will imply functional identities for perturbed defect operators via the relation described in Section \ref{sec:def}. The same holds for left duals.
\\[.3em]
(iii) The functors $\mathcal{R}_\varphi$ defined in Remark \ref{rem:CF} are compatible with these dualities in the sense that $\mathcal{R}_\varphi( (U_f)^\vee ) = (\mathcal{R}_\varphi(U_f))^\vee$ and $\mathcal{R}_\varphi( \eev(U_f) ) = \eev(\mathcal{R}_\varphi(U_f))$.
This follows from $c(f \circ (\varphi \otimes \id_U)) = c(f) \circ (\varphi \otimes \id_{U^\vee})$ and
$\tilde c(f \circ (\varphi \otimes \id_U)) = \tilde c(f) \circ (\varphi \otimes \id_{\eev U})$.
\end{Rem}

\section{Relation to defect operators}\label{sec:def}

\subsection{Topological defect lines}\label{sec:top-def}

Defects are lines on the world sheet where the fields can be discontinuous or even singular. Suppose we are given a CFT that is well-defined on surfaces with defect lines, that is, it satisfies the axioms in \cite[Sect.\,3]{Runkel:2008gr} (or at least a genus 0 version thereof). To a defect we can assign a linear operator $D$ on the space of states $\Hc$ of the CFT. This operator can be extracted by wrapping the defect line around a short cylinder $[-\eps,\eps] \times S^1$, where we place two states $u$ and $v$ on the two boundary circles. The resulting amplitude, in the limit $\eps \rightarrow 0$, is the pairing $\langle u, D v \rangle$.

Working with fields rather than with states, the defect operator $D$ is obtained as the correlator assigned to the Riemann sphere $\Cb \cup \{\infty\}$ with one in-going puncture at $0$ and one out-going puncture at $\infty$, both with standard local coordinates, and a defect line placed on the unit circle $S^1$. By the state-field correspondence, the space of states $\Hc$ is at the same time the space of local bulk fields, so that again $D : \Hc \rightarrow \Hc$.

We call the defect {\em conformal} if it is transparent to the field $T-\bar T$, the difference of the holomorphic and anti-holomorphic parts of the stress tensor. In terms of modes, this condition reads
\be
  \text{$D$ conformal} ~~ \Leftrightarrow ~~ [\,L_m-\bar L_{-m}\,,\,D\,] = 0 ~.
\ee
This includes totally transmitting defects, such as the invisible defect described by the identity operator $D=\id$, and totally reflecting defects, such as the product of two boundary states $D = |a\rangle\!\rangle \langle\!\langle b|$. Here we are interested in the totally transmitting case, more precisely in defects which are transparent to both $T$ and $\bar T$ separately. Such defects were first investigated in the context of rational CFT in \cite{Petkova:2000ip} and were termed {\em topological defects} in \cite{Bachas:2004sy},
\be
  \text{$D$ topological} ~~ \Leftrightarrow ~~ [L_m,D] = 0 = [\bar L_m,D] ~.
\labl{eq:top-def-cond}

We will be working in rational CFT, so that the chiral algebra of the CFT will be a rational vertex operator algebra $V$ (recall footnote \ref{fn:rational}). Denote by $\Cc = \Rep(V)$ the category of (appropriate) representations of $V$. It is a semi-simple finite rigid braided monoidal category which is modular \cite{hl-vtc,Huang2005} (see \cite[Sect.\,3.1]{BaKi-book} for more on modular categories). We will not need many details about modular categories, but we note that $\Cc$ satisfies the conditions of Theorems \ref{thm:CF-tens} and \ref{thm:CF-rigid}.

Let us pick a set of representatives\footnote{\label{fn:Ri-vs-Rf-notation}
  The notation $R_i$, where $i$ is an index of a simple object, should not be confused with the notation $R_f$ for objects of $\mathcal{C}_F$ (for some $F$), where $f : F \otimes R \rightarrow R$ is a morphism. The meaning of the index should be clear from the context, and in any case we will mostly use $i,j,k$ for indices of simple objects and $f,g,h$, as well as $c$ and $x$, for morphisms.}
$\{\,R_i\,|\,i \in \Ic\,\}$ of the isomorphism classes of simple objects\footnote{An object $U$ is simple iff it does not have proper subobjects, that is, iff every monomorphism $s:S\rightarrow U$ is either zero or an isomorphism.},
so that $\one \equiv R_0 \equiv V$ is the monoidal unit.
We restrict ourselves in this paper to the Cardy case constructed from $V$. The space of states of this model is
\be
  \Hc = \bigoplus_{i \in\Ic} R_i \otimes_\Cb R_i^\vee ~,
\ee
where $R_i^\vee$ denotes the contragredient representation to $R_i$.
Also, we will only consider topological defects which are maximally symmetric in that they are compatible with the entire chiral symmetry $V \otimes_\Cb V$, i.e.\ \eqref{eq:top-def-cond} holds for the modes of all fields in $V \otimes_\Cb V$ not just for those of the stress tensor. According to \cite{Petkova:2000ip,tft1} the different maximally symmetric topological defects are labelled by representations of $V$, that is, objects $R \in \Cc$. We denote the defect operator of the defect labelled by $R \in \Cc$ by $D[R]$. The defect operator assigned to a simple object $R_i$ is \cite{Petkova:2001ag,defect}
\be
  D[R_i] = \sum_{j \in \Ic} \frac{S_{ij}}{S_{0j}} \id_{R_j \otimes_\Cb R_j^\vee} ~,
\labl{eq:top-def-op}
where by $\id_{R_j \otimes_\Cb R_j^\vee}$ we mean the projector to the direct summand $R_j \otimes_\Cb R_j^\vee$ of $\Hc$, and $S$ is the modular matrix, i.e.\ the $|\Ic|{\times}|\Ic|$-matrix which describes the modular transformation of characters. If $R \cong \bigoplus_{i \in \Ic} (R_i)^{\oplus n_i}$ then $D[R] = \sum_{i \in \Ic} n_i\,D[R_i]$.

\subsection{Correlators of chiral defect fields}\label{sec:corr-def}

By a {\em chiral defect field} we mean a field that `lives on the defect' and that has left/right conformal weight $(h,0)$. The notion of defect fields is described for example in \cite[Sect.\,3.4]{tft4} and \cite[Sect.\,3.2]{Runkel:2008gr}. The defect fields have well-defined weights with respect to $L_0$ and $\bar L_0$ because we are considering topological defects, and those are transparent to the holomorphic and anti-holomorphic part of the stress tensor.

The space of chiral defect fields on a defect labelled by $R \in \Cc$ consists of all vectors $v \otimes_\Cb \Omega \in (R \otimes R^\vee) \otimes_\Cb V$, where $\Omega \in V$ is the vacuum vector of $V$
and the tensor product $R \otimes R^\vee$ is the fusion tensor product in $\Cc$, see \cite[Eqn.\,(3.37)]{tft4} and \cite{Petkova:2000ip,Petkova:2001ag,tft1}. Pick a representation $F \in \Cc$. A {\em chiral defect field in representation} $F$ is specified by a vector $\phi \in F$ and a morphism $\tilde f : F \rightarrow R \otimes R^\vee$ in $\Cc$. Instead of $\tilde f$ we find it more convenient to give a morphism $f : F \otimes R \rightarrow R$.

We are going to define a defect operator for a defect labelled by a representation $R$ with chiral defect fields $\phi$ inserted at mutually distinct points $e^{i \theta_1},\dots,e^{i \theta_n}$ on the unit circle, where for each insertion we allow a different morphism $f_1,\dots,f_n$. We will denote this operator by
\be
  D[R;f_1,\dots,f_n;\theta_1,\dots,\theta_n] : \Hc \rightarrow \overline\Hc ~.
\labl{eq:DR-thetas}
The operator $D$ may have contributions in an infinite number of graded components of the target vector spaces. Hence we have to pass to a completion of $\mathcal{H}$, namely to the direct product $\overline\Hc$ of the graded components of $\mathcal{H}$. We will later integrate over the variables $\theta_k$, and the resulting operator commutes with the grading, so that we obtain an operator $\Hc \rightarrow \Hc$.

Let us restrict $D$ to the sector $R_i \otimes_\Cb R_i^\vee$ of $\Hc$ and call the resulting operator $D_i$. Because the defect fields are all chiral, they do not affect the anti-holomorphic sector, and hence the image of $D_i$ will lie entirely in the summand $\overline{R_i \otimes_\Cb R_i^\vee}$ of $\overline\Hc$. The operator $D_i$ is an element of a suitable space of conformal blocks, namely of a tensor product (over $\Cb$) of two spaces of conformal blocks on the sphere, related to the two chiral halfs of the CFT. On the first sphere $\Cb \cup \{\infty\}$ we have insertions of $R_i$ at $0$ and $\infty$, and of $F$ at $e^{i \theta_1},\dots,e^{i \theta_n}$. Insertions at $\infty$ will always be treated as out-going, the others as in-going. Because the defect fields are chiral, on the second sphere we just have insertions of $R_i^\vee$ at $0$ and $\infty$. Altogether, the conformal block is an operator
\be
  \mathscr{C}[R;f_1,\dots,f_n;\theta_1,\dots,\theta_n]_i : R_i \otimes_\Cb R_i^\vee \otimes_\Cb F \otimes_\Cb \cdots \otimes_\Cb
  F \longrightarrow \overline{R_i \otimes_\Cb R_i^\vee} ~.
\labl{eq:def-C-operator}
It determines the defect operator $D_i$ on a vector $u \otimes v \in R_i \otimes_\Cb R_i^\vee \subset \Hc$ via
\be
  D_i(u \otimes v)
  =  \mathscr{C}[R;f_1,\dots,f_n;\theta_1,\dots,\theta_n]_i(u \otimes v \otimes \phi \otimes \cdots \otimes \phi) ~.
\labl{eq:def-D-on-vect}
The vector space of conformal blocks from which \eqref{eq:def-C-operator} is taken is finite-dimensional, as is always the case in rational CFT, but its dimension can be 
quite high and will grow with the number $n$ of insertions. We thus need an efficient method to specify elements in the space of conformal blocks. Such a method is provided by using three-dimensional topological field theory to describe correlators of rational CFT, see \cite{Felder:1999mq,tft1} and also \cite{tft4,Frohlich:2004ef,defect}, which treat defect lines and defect fields in detail.

The 3d TFT assigns to a three-manifold $M$ with embedded framed Wilson graph (to be called a 
ribbon graph) an element in the space of conformal blocks on the boundary surface $\partial M$ of $M$. If the 3d TFT is Chern-Simons theory for a gauge group $G$, the conformal blocks are those of the corresponding WZW model \cite{Witten:1988hf,Frohlich:1989gr}. There is also a general construction, whereby the 3d TFT is defined by a modular category $\Cc$ \cite{Turaev-book,BaKi-book}, which in turn is obtained from the representations of a rational vertex operator algebra \cite{Moore:1989vd,Huang2005}. Let us denote this TFT as $\mathrm{tft}_\Cc$.

In the TFT approach to correlators of rational CFT, one starts from a world sheet $X$, possibly with boundary and defect lines, and with various field insertions, and constructs from 
this a three-manifold $M_X$ with embedded ribbon graph. The boundary of $M_X$ is the double $\hat X$ of the surface $X$ and the TFT assigns to $M_X$ a conformal block in $\hat X$, which we write as $\mathrm{tft}_\Cc(M_X)$. This is the correlator for the world sheet $X$.

Let us see how this works in the case at hand, where $X$ is $\Cb \cup \{ \infty \}$ with bulk fields in representation $R_i \otimes_\Cb R_i^\vee$ inserted at $0$ and $\infty$, and with a defect line labelled $R$ placed on the unit circle on which defect fields in representation $F$ are inserted at the points $e^{i \theta_1},\dots,e^{i \theta_n}$. As $X$ is oriented and has empty boundary, the three-manifold is simply $M_X = X \times [-1,1]$. Note that $\partial M_X$ does indeed consist of two Riemann spheres, so that the TFT will determine an element in the tensor product of two spaces of conformal blocks on the sphere, as discussed above. It remains to construct the ribbon graph embedded in $M_X$. To do this, we place a circular ribbon labelled by the representation $R$ on the unit circle in the plane $X \times \{0\}$. This ribbon is connected to the marked points $e^{i \theta_k}$ on the boundary $X \times \{1\}$ of $M_X$ with ribbons labelled by $F$. The junction of $F$ and $R$ is formed by the intertwiner $f_k : F \otimes R \rightarrow R$. For the bulk insertions at $0$ and $\infty$ one places a vertical ribbon inside $M_X$ connecting the marked points on the boundary components $X \times \{1\}$ and $X \times \{-1\}$. The resulting ribbon graph is
\bea
  M[R;f_1,\dots,f_n;\theta_1,\dots,\theta_n]_i =
\\[.3em]
      \raisebox{-100pt}{
  \begin{picture}(450,220)
   \put(0,3){\scalebox{.90}{\includegraphics{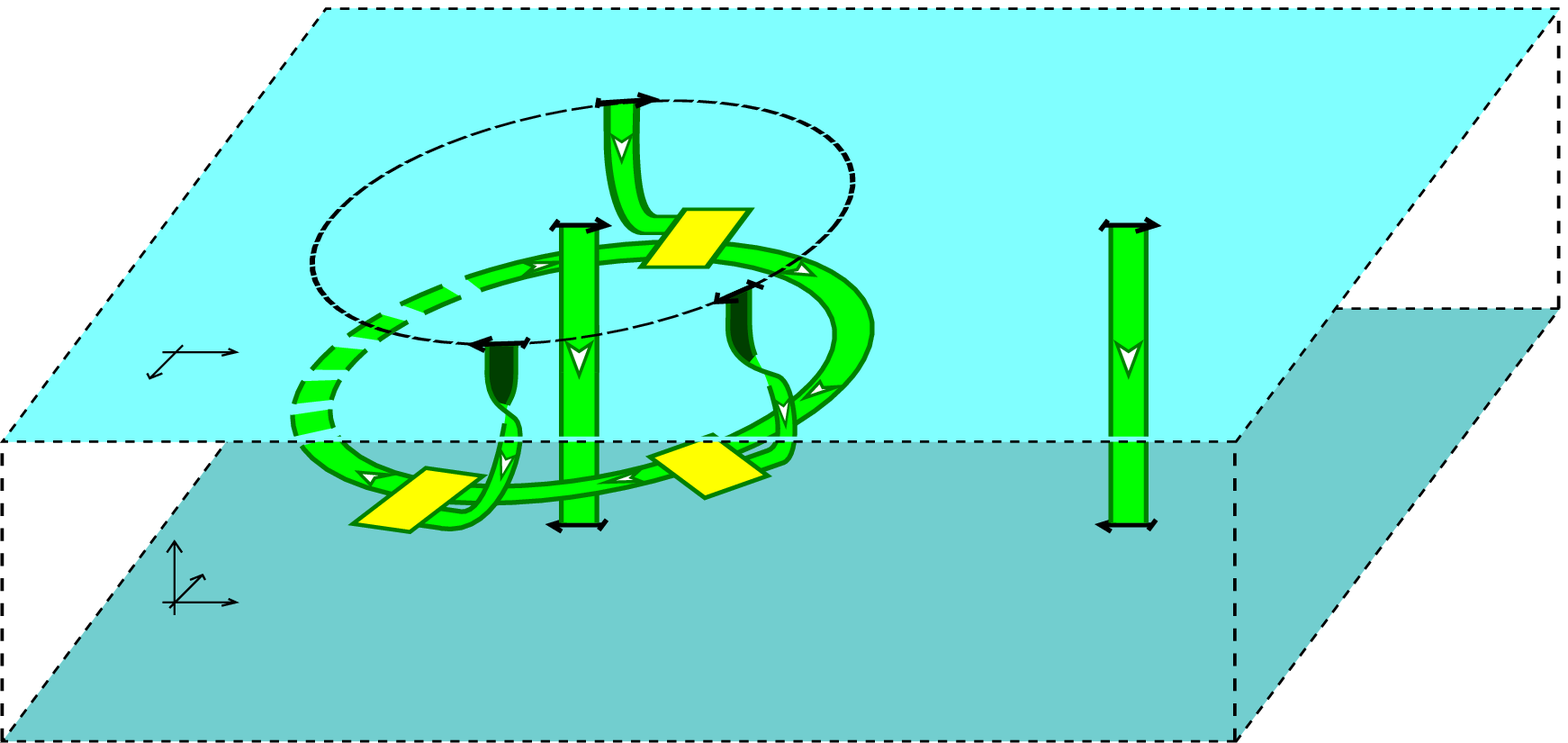}}}
   \put(0,3){
     \setlength{\unitlength}{.90pt}\put(-1,-1){
     \put( 54,140) {\scriptsize$ \Cb \cup \{ \infty \} $}
     \put(-12,94) {\scriptsize$ -1 $}
     \put(-7,47) {\scriptsize$ 0 $}
     \put(-7,0) {\scriptsize$ 1 $}
     \put( 78, 44) {\scriptsize$ 1 $}
     \put( 67, 56) {\scriptsize$ 2 $}
     \put( 58, 65) {\scriptsize$ 3 $}
     \put( 78,122) {\scriptsize$ 1 $}
     \put( 45,110) {\scriptsize$ 2 $}
     \put(176,172) {\scriptsize$ z{=}0 $}
     \put(354,172) {\scriptsize$ z{=}\infty $}
     \put(194,110) {\scriptsize$ R_i $}
     \put(372,110) {\scriptsize$ R_i^\vee $}
     \put(222, 88) {\scriptsize$ f_{\sigma 1} $}
     \put(130, 78) {\scriptsize$ f_{\!\sigma\!2} $}
     \put(215,160) {\scriptsize$ f_{\sigma n} $}
     \put(228,148) {\scriptsize$ \theta_{\!\sigma\!1} $}
     \put(156,134) {\scriptsize$ \theta_{\sigma 2} $}
     \put(191,212) {\scriptsize$ \theta_{\sigma n} $}
     \put(226,128) {\scriptsize$ F $}
     \put(147,116) {\scriptsize$ F $}
     \put(186,186) {\scriptsize$ F $}
     \put(282,127) {\scriptsize$ R $}
     \put(205, 76) {\scriptsize$ R $}
     \put( 96, 81) {\scriptsize$ R $}
     \put(154,156) {\scriptsize$ R $}
     }\setlength{\unitlength}{1pt}}
  \end{picture}}
\eear
\labl{eq:def-M-def}
For the TFT conventions used here, see \cite[Sect.\,2]{tft1}, and for more details on the construction of the ribbon graph consult \cite[Sect.\,3\,\&\,4]{tft4}. The orientation of the `top' plane of $M$ is obtained from that of $M$ by taking the inward pointing normal.
The arrows at the ends of the ribbons refer to a particular choice of local coordinates around the $F$-insertions, namely the local coordinate at $\exp(i \theta_{\sigma k})$ is given by $\zeta \mapsto -i (\exp(-i \theta_{\sigma k}) \zeta-1)$, so that $\exp(i \theta_{\sigma k})$ gets mapped to zero and the real axis of the local coordinate system is tangent to the defect circle.
We do not demand that the $\theta_1,\dots,\theta_n$ are ordered. Instead we define $\sigma \in S_n$ to be the unique permutation of $n$ elements for which $0 \le \theta_{\sigma 1} < \theta_{\sigma 2} \cdots < \theta_{\sigma n} < 2 \pi$. Finally, the conformal block \eqref{eq:def-C-operator} is given by
\be
  \mathscr{C}[R;f_1,\dots,f_n;\theta_1,\dots,\theta_n]_i
  = \mathrm{tft}_\Cc\big(M[R;f_1,\dots,f_n;\theta_1,\dots,\theta_n]_i  \big) ~.
\labl{eq:def-C-via-M}
One can work out this conformal block in terms of intertwiners as in \cite[Sect.\,5]{tft4}, but we will not need such an explicit expression here. This conformal block in turn determines the defect operator \eqref{eq:DR-thetas} via $D = \bigoplus_i D_i$ with $D_i$ given in \eqref{eq:def-D-on-vect}.

The strength of the representation \eqref{eq:def-C-via-M} lies in the fact that we can now use identities that hold within the 3d TFT, i.e.\ manipulations which change the ribbon graph inside $M$ without modifying the value of $\mathrm{tft}_\Cc(M)$, to prove identities among conformal blocks. This will be used extensively in the proof of the next lemma. In fact, the manipulations below will only involve a neighbourhood of the circular ribbon in \eqref{eq:def-M-def}. For this reason, it is convenient to have a shorthand for \eqref{eq:def-M-def} which only shows this region of $M$. We will write
\be
  M[R;f_1,\dots,f_n;\theta_1,\dots,\theta_n]_i = M\!\!\left[
    \raisebox{-18pt}{
  \begin{picture}(188,45)
   \put(0,3){\scalebox{.90}{\includegraphics{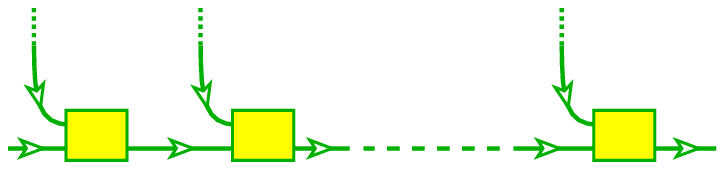}}}
   \put(0,3){
     \setlength{\unitlength}{.90pt}\put(-4,-4){
     \put( 24, 9) {\scriptsize$ f_{\sigma 1} $}
     \put( 72, 9) {\scriptsize$ f_{\sigma 2} $}
     \put(176, 9) {\scriptsize$ f_{\sigma n} $}
     \put( 15,40) {\scriptsize$ \theta_{\sigma 1} $}
     \put( 63,40) {\scriptsize$ \theta_{\sigma 2} $}
     \put(167,40) {\scriptsize$ \theta_{\sigma n} $}
     \put( 3,27) {\scriptsize$ F $}
     \put(51,27) {\scriptsize$ F $}
     \put(155,27) {\scriptsize$ F $}
     \put(  7,-3) {\scriptsize$ R $}
     \put( 47,-3) {\scriptsize$ R $}
     \put( 92,-3) {\scriptsize$ R $}
     \put(161,-3) {\scriptsize$ R $}
     \put(196,-3) {\scriptsize$ R $}
     }\setlength{\unitlength}{1pt}}
  \end{picture}}
  \right] ~.
\labl{eq:def-M-short}

\begin{Lem}\label{lem:D-prop}
(i) Let $0 \rightarrow K_h \rightarrow R_f \rightarrow C_c \rightarrow 0$ be an exact sequence in $\Cc_F$, and let $\theta_1,\dots,\theta_m \in [0,2 \pi[$ be mutually distinct. Then
\be
  D[R;f,\dots,f;\theta_1,\dots,\theta_m]
  = D[K;h,\dots,h;\theta_1,\dots,\theta_m] + D[C;c,\dots,c;\theta_1,\dots,\theta_m]
\ee
(ii) Let $R_f, S_g \in \Cc_F$, and let $\theta_1,\dots,\theta_m,\eta_1,\dots,\eta_n \in [0,2 \pi[$ be mutually distinct. Then
\be
 \begin{array}{l}\displaystyle
  \lim_{\eps \rightarrow 0+} D[R;f,\dots,f;\theta_1,\dots,\theta_m] ~ e^{\eps(L_0+\bar L_0)} ~ D[S;g,\dots,g;\eta_1,\dots,\eta_n]
  \\ \displaystyle \qquad
  =~ D[R \otimes S; T(f,0),\dots,T(f,0),T(0,g),\dots,T(0,g);\theta_1,\dots,\theta_m,\eta_1,\dots,\eta_n]
  \end{array}
\labl{eq:D-prop-ii}
\end{Lem}

\begin{proof}
(i) Denote the morphisms in the exact sequence by $e_K : K_h \rightarrow R_f$ and $r_C : R_f \rightarrow C_c$. In the present situation, the category $\Cc = \Rep(V)$ is modular, and thus in particular semi-simple. Therefore, in $\Cc$ the exact sequence $0 \rightarrow K \xrightarrow{e_K} R \xrightarrow{r_C} C \rightarrow 0$ splits, i.e.\ we can find $r_K : R \rightarrow K$ and $e_C : C \rightarrow R$ such that $r_K \circ e_K = \id_K$, $r_C \circ e_C = \id_C$, and $e_K \circ r_K + e_C \circ r_C = \id_R$. Using the decomposition of $\id_R$ we can write
\be
  \mathscr{C}[R;f,\dots,f;\theta_1,\dots,\theta_n]_i = \mathrm{tft}_\Cc(M_K) + \mathrm{tft}_\Cc(M_C)
\labl{eq:D-prop-pf1}
where
\bea
  M_K = M\!\left[
    \raisebox{-18pt}{
  \begin{picture}(230,45)
   \put(0,3){\scalebox{.90}{\includegraphics{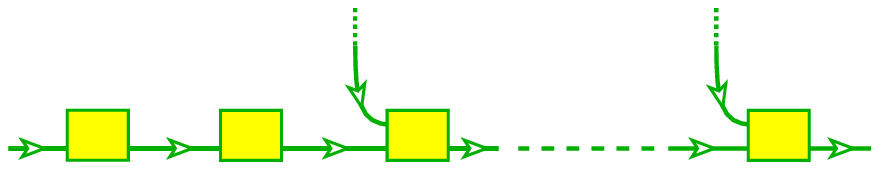}}}
   \put(0,3){
     \setlength{\unitlength}{.90pt}\put(-6,-6){
     \put( 28, 13) {\scriptsize$ r_K $}
     \put( 72, 13) {\scriptsize$ e_K $}
     \put(122, 13) {\scriptsize$ f $}
     \put(225, 13) {\scriptsize$ f $}
     \put( 110,40) {\scriptsize$ \theta_{\sigma 1} $}
     \put(215,40) {\scriptsize$ \theta_{\sigma n} $}
     \put(98,27) {\scriptsize$ F $}
     \put(202,27) {\scriptsize$ F $}
     \put(10,-1) {\scriptsize$ R $}
     \put( 53,-1) {\scriptsize$ K $}
     \put( 98,-1) {\scriptsize$ R $}
     \put( 138,-1) {\scriptsize$ R $}
     \put(202,-1) {\scriptsize$ R $}
     \put(245,-1) {\scriptsize$ R $}
     }\setlength{\unitlength}{1pt}}
  \end{picture}}
  \right] ~~,
  \\[3em]\displaystyle
  M_C = M\!\!\left[
    \raisebox{-18pt}{
  \begin{picture}(230,45)
   \put(0,3){\scalebox{.90}{\includegraphics{pic03.eps}}}
   \put(0,3){
     \setlength{\unitlength}{.90pt}\put(-6,-6){
     \put( 28, 13) {\scriptsize$ r_C $}
     \put( 72, 13) {\scriptsize$ e_C $}
     \put(122, 13) {\scriptsize$ f $}
     \put(225, 13) {\scriptsize$ f $}
     \put( 110,40) {\scriptsize$ \theta_{\sigma 1} $}
     \put(215,40) {\scriptsize$ \theta_{\sigma n} $}
     \put(98,27) {\scriptsize$ F $}
     \put(202,27) {\scriptsize$ F $}
     \put(10,-1) {\scriptsize$ R $}
     \put( 53,-1) {\scriptsize$ C $}
     \put( 98,-1) {\scriptsize$ R $}
     \put( 138,-1) {\scriptsize$ R $}
     \put(202,-1) {\scriptsize$ R $}
     \put(245,-1) {\scriptsize$ R $}
     }\setlength{\unitlength}{1pt}}
  \end{picture}}
  \right]
~~.
\eear
\ee
Since $e_K : K_h \rightarrow R_f$ is a morphism in $\Cc_F$, it satisfies the identity $e_K \circ h = f \circ (\id_F \otimes e_K)$. This can be used to move $e_K$ past $f$, for example,
\be
  \mathrm{tft}_\Cc(M_K) = \mathrm{tft}_\Cc\!\left(M\!\!\left[
    \raisebox{-18pt}{
  \begin{picture}(230,45)
   \put(0,3){\scalebox{.90}{\includegraphics{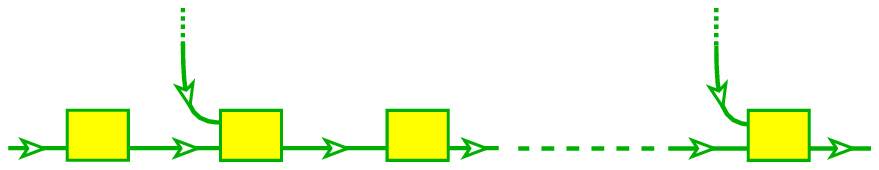}}}
   \put(0,3){
     \setlength{\unitlength}{.90pt}\put(-6,-6){
     \put( 28, 12) {\scriptsize$ r_K $}
     \put( 73, 12) {\scriptsize$ h $}
     \put(120, 12) {\scriptsize$ e_K $}
     \put(225, 12) {\scriptsize$ f $}
     \put(60,40) {\scriptsize$ \theta_{\sigma 1} $}
     \put(215,40) {\scriptsize$ \theta_{\sigma n} $}
     \put(48,27) {\scriptsize$ F $}
     \put(202,27) {\scriptsize$ F $}
     \put(10,-1) {\scriptsize$ R $}
     \put( 53,-1) {\scriptsize$ K $}
     \put( 98,-1) {\scriptsize$ K $}
     \put( 138,-1) {\scriptsize$ R $}
     \put(202,-1) {\scriptsize$ R $}
     \put(245,-1) {\scriptsize$ R $}
     }\setlength{\unitlength}{1pt}}
  \end{picture}}
  \right]\right)~~.
\ee
If one repeats this procedure and in this way takes $e_K$ around the loop, one arrives at
\bea
  \mathrm{tft}_\Cc(M_K) = \mathrm{tft}_\Cc\!\left(M\!\!\left[
    \raisebox{-18pt}{
  \begin{picture}(230,45)
   \put(0,3){\scalebox{.90}{\includegraphics{pic03.eps}}}
   \put(0,3){
     \setlength{\unitlength}{.90pt}\put(-6,-6){
     \put( 28, 12) {\scriptsize$ e_K $}
     \put( 72, 12) {\scriptsize$ r_K $}
     \put(122, 12) {\scriptsize$ h $}
     \put(225, 12) {\scriptsize$ h $}
     \put( 110,40) {\scriptsize$ \theta_{\sigma 1} $}
     \put(215,40) {\scriptsize$ \theta_{\sigma n} $}
     \put(98,27) {\scriptsize$ F $}
     \put(202,27) {\scriptsize$ F $}
     \put(10,-1) {\scriptsize$ K $}
     \put( 53,-1) {\scriptsize$ R $}
     \put( 98,-1) {\scriptsize$ K $}
     \put( 138,-1) {\scriptsize$ K $}
     \put(202,-1) {\scriptsize$ K $}
     \put(245,-1) {\scriptsize$ K $}
     }\setlength{\unitlength}{1pt}}
  \end{picture}}
  \right]\right)
\\[2.5em]\displaystyle
 = \mathscr{C}[K;h,\dots,h;\theta_1,\dots,\theta_n]_i ~.
\eear
\labl{eq:D-prop-pf2}
In the last step we used $r_K \circ e_K = \id_K$ and Equation \eqref{eq:def-C-via-M}. For $\mathrm{tft}_\Cc(M_C)$ one proceeds similarly, only that here $r_C : R_f \rightarrow C_c$ is the morphism in $\Cc_F$, and so one has to move $r_C$ around the loop in the opposite sense. This results in
\be
  \mathrm{tft}_\Cc(M_C) = \mathscr{C}[C;c,\dots,c;\theta_1,\dots,\theta_n]_i ~.
\labl{eq:D-prop-pf3}
Combining \eqref{eq:D-prop-pf1}, \eqref{eq:D-prop-pf2} and  \eqref{eq:D-prop-pf3} establishes part (i) of the lemma.
\\[.3em]
(ii) Because the conformal block in \eqref{eq:def-C-via-M} is a map from $R_i \otimes_\Cb R_i^\vee$ to the direct product $\overline{R_i \otimes_\Cb R_i^\vee}$ of the $L_0$,$\Lb_0$-eigenspaces in $R_i \otimes_\Cb R_i^\vee$, we have to take care that the composition is well-defined. This is ensured by the exponential in \eqref{eq:D-prop-ii}. Since the insertion points $e^{i \theta}$ of the intertwining operators (of the vertex operator algebra representations) are distinct, the limit $\eps\rightarrow 0$ is well-defined. Let $\mathscr{C}_\text{lhs}$ and $\mathscr{C}_\text{rhs}$ be the conformal blocks obtained from the left and right hand side of \eqref{eq:D-prop-ii}, respectively. To see that $\mathscr{C}_\text{lhs}=\mathscr{C}_\text{rhs}$ we again use the 3d\,TFT. Let us look at a particular example of the ordering of the $\theta_k$ and $\eta_k$, say $\theta_1<\eta_1<\eta_2<\theta_2<\dots<\eta_n<\theta_m$. The general case works along the same lines. Substituting the definitions, one finds that the three-manifold and ribbon graph for $\mathscr{C}_\text{rhs}$ is
\be
  \mathscr{C}_\text{lhs} = \mathscr{C}_\text{rhs} = \mathrm{tft}_\Cc\!\left(\!M\!\!\left[\!\!
    \raisebox{-30pt}{
  \begin{picture}(300,68)
   \put(0,3){\scalebox{.90}{\includegraphics{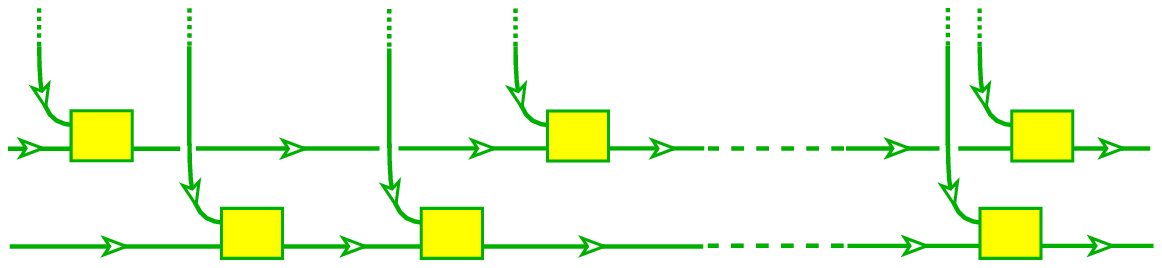}}}
   \put(0,3){
     \setlength{\unitlength}{.90pt}\put(-5,-8){
     \put( 72, 15) {\scriptsize$ g_{1} $}
     \put(129, 15) {\scriptsize$ g_{2} $}
     \put(290, 15) {\scriptsize$ g_{n} $}
     \put(28,42) {\scriptsize$ f_{1} $}
     \put(166,42) {\scriptsize$ f_{2} $}
     \put(300,42) {\scriptsize$ f_{n} $}
     \put(46,75) {\scriptsize$ \eta_{1} $}
     \put(103,75) {\scriptsize$ \eta_{2} $}
     \put(265,75) {\scriptsize$ \eta_{n} $}
     \put(19,75) {\scriptsize$ \theta_{1} $}
     \put(156,75) {\scriptsize$ \theta_{2} $}
     \put(291,75) {\scriptsize$ \theta_{m} $}
     \put(48,27) {\scriptsize$ F $}
     \put(105,27) {\scriptsize$ F $}
     \put(267,27) {\scriptsize$ F $}
     \put(5,62) {\scriptsize$ F $}
     \put(143,62) {\scriptsize$ F $}
     \put(289,62) {\scriptsize$ F $}
     \put( 33,1) {\scriptsize$ S $}
     \put( 103,1) {\scriptsize$ S $}
     \put( 172,1) {\scriptsize$ S $}
     \put(265,1) {\scriptsize$ S $}
     \put(320,1) {\scriptsize$ S $}
     \put( 84,47) {\scriptsize$ R $}
     \put( 138,47) {\scriptsize$ R $}
     \put( 190,47) {\scriptsize$ R $}
     \put(258,47) {\scriptsize$ R $}
     \put(320,47) {\scriptsize$ R $}
     }\setlength{\unitlength}{1pt}}
  \end{picture}}
  \!\right]
\right)  ~.
\labl{eq:def-com-aux1}
To see that $\mathscr{C}_\text{lhs}$ leads to the same result, one has to translate the composition of conformal blocks into a gluing of three-manifolds as in \cite[Thm.\,3.2]{Felder:1999mq}. Namely, one needs to cut out a cylinder around the $R_i$-ribbon at $z=0$ of $D[R;\dots]$ and around the $R_i^\vee$-ribbon at $z=\infty$ of $D[S;\dots]$, and identify the resulting cylindrical boundaries. The resulting ribbon graph can be deformed to give \eqref{eq:def-com-aux1}. This establishes part (ii) of the lemma.
\end{proof}

\subsection{Perturbed topological defects}\label{sec:pert-def}

The operator of the perturbed defect is defined via an exponentiated integral. That is, for an object $R_f \in \Cc_F$ we set\footnote{
  Recall from below \eqref{eq:def-C-via-M} that the local coordinate around the insertion of a defect field $\phi$ at $e^{i \theta}$ was chosen to be $\zeta \mapsto -i(e^{-i \theta} \zeta - 1)$. This choice makes (for example) $D[R;f;\theta]$ periodic under $\theta \leadsto \theta+2\pi$. Had we instead chosen the standard local coordinates $\zeta \mapsto \zeta-p$ on the complex plane around a point $p$, $D[R;f;\theta]$ would have picked up the phase $e^{-2 \pi i h_\phi}$.}
\be
  D[R_f] = \sum_{n=0}^\infty \frac{1}{n!}\, D[R_f]^{(n)}
  \quad , \quad
  D[R_f]^{(n)}
  = \int_{0}^{2\pi} \mem D[R;f,\dots,f;\theta_1,\dots,\theta_n] \, d\theta_1 \cdots d\theta_n ~.
\labl{eq:D[Rf]-def}
Because of the permutation that orders the arguments in the definition \eqref{eq:def-M-def}, \eqref{eq:def-C-via-M} and \eqref{eq:def-D-on-vect} of the defect operator, a path-ordering prescription is automatically imposed and does not need to be included explicitly in the integration regions for $D[R_f]^{(n)}$.
The integrals in $D[R_f]^{(n)}$ and the infinite sum in $D[R_f]$ may or may not converge. In lack of a direct way to ensure convergence, we say that an object $R_f \in \Cc_F$ {\em has finite integrals} if $\varphi(D[R_f]^{(n)}v)$ exists for each $\varphi \in \Hc^*$, $v\in \Hc$, and $n \ge 0$. Note that this is not a property of the category $\Cc_F$ alone, but instead also depends on the vertex operator algebra $V$ and the vector $\phi \in F$. Generically one expects that if the element $\phi \in F$ has conformal weight $h_\phi <\tfrac12$, then all $R_f \in \Cc_F$ have finite integrals (but we have no proof). Let $R_f \in \Cc_F$ have finite integrals. It is demonstrated in \cite[Sect.\,2.2]{Runkel:2007wd} that
\be
   [L_0, D[R_f]^{(n)}] = 0
   \quad \text{and} \quad
   [\Lb_m, D[R_f]^{(n)}] = 0 ~ \text{for all} ~ m \in \Zb.
\ee
We will not discuss the convergence of the infinite sum in \eqref{eq:D[Rf]-def}.
Instead we will treat it as a formal power series in the following way. For $\zeta \in \Cb$ we have $D[R_{\zeta f}]^{(n)} = \zeta^n \, D[R_f]^{(n)}$. Now take $\zeta$ to be a formal parameter and let us define, by slight abuse of notation,
\be
  D[R_{\zeta f}] = \sum_{n=0}^\infty \frac{\zeta^n}{n!}\, D[R_f]^{(n)} ~ \in \End(\Hc)[\![\zeta]\!] ~.
\ee

\begin{Thm}\label{thm:D-prop}
Let $\zeta$ be a formal parameter.\\
(i) Let $0 \rightarrow K_h \rightarrow R_f \rightarrow C_c \rightarrow 0$ be an exact sequence in $\Cc_F$, and let $K_h$, $R_f$, $C_c$ have finite integrals. Then $D[R_{\zeta f}] = D[K_{\zeta h}] + D[C_{\zeta c}]$.
\\
(ii) Let $R_f, S_g \in \Cc_F$ have finite integrals. Then $D[R_{\zeta f}]D[S_{\zeta g}]  = D[(R\otimes S, \zeta T(f,g))]$.
\end{Thm}

\begin{proof}
Part (i) holds because by Lemma \ref{lem:D-prop}\,(i) it already holds before integration.
For part (ii) first note that the exponential in \eqref{eq:D-prop-ii} is not necessary to make the composition $D[R_{\zeta f}]D[S_{\zeta g}]$ well-defined, because $D[R_{\zeta f}]$ commutes with $L_0+\Lb_0$ and we can write $D[R_{\zeta f}]D[S_{\zeta g}] = \lim_{\eps \rightarrow 0}  e^{-\eps(L_0+\bar L_0)} D[R_{\zeta f}] e^{\eps(L_0+\bar L_0)} D[S_{\zeta g}]$. We will therefore not write the limit in the equations below.
Define operators $A_n$ and $B_n$ via
\be
  D[R_{\zeta f}]D[S_{\zeta g}] = \sum_{n>0}^\infty \frac{1}{n!} \zeta^n A_n \quad \text{and} \quad
  D[(R\otimes S, \zeta T(f,g))] = \sum_{n>0}^\infty \frac{1}{n!} \zeta^n B_n ~.
\ee
We have to show that $A_n=B_n$. Starting from $A_n$ we find
\bea
  A_n = \sum_{m=0}^n \binom{n}{m} D[R_{\zeta f}]^{(m)} D[S_{\zeta g}]^{(n-m)}
  \enl
  = \sum_{m=0}^n \binom{n}{m} \int
  D[R;f,\dots,f;\theta_1,\dots,\theta_m] \, D[S;g,\dots,g;\eta_1,\dots,\eta_{n-m}]
  \enl
  = \sum_{m=0}^n \binom{n}{m} \int
  D[R \otimes S; T(f,0),\dots,T(f,0),T(0,g),\dots,T(0,g);\theta_1,\dots,\theta_m,\eta_1,\dots,\eta_{n-m}]
\eear\labl{eq:D-thm-pf1}
where $\int \equiv \int_0^{2 \pi} d\theta_1\cdots d\theta_m d\eta_1\cdots d\eta_{n-m}$ and in the last step we used Lemma \ref{lem:D-prop}\,(ii). For $B_n$ we get
\be
  B_n = \int_0^{2\pi} \mem d\alpha_1 \cdots d\alpha_n ~
  D[R \otimes S; T(f,g),\dots,T(f,g);\alpha_1,\dots,\alpha_n]~.
\ee
To see that this is equal to the right hand side of \eqref{eq:D-thm-pf1} one first writes $T(f,g) = T(f,0) + T(0,g)$, then expands out the integrand into $2^n$ summands and groups together those with the same number of $T(f,0)$ and $T(0,g)$. The distinct ordering in each term can be absorbed into a change of integration variables as the angles $\alpha_k$ are all integrated from $0$ to $2\pi$.
\end{proof}

Theorem \ref{thm:D-prop} implies the following corollary.

\begin{Cor}\label{cor:rep-Gr}
Let $\zeta$ be a formal parameter and let $R_f, S_g \in \Cc_F$ have finite integrals.\\
(i) If $[R_f] = [S_g]$ in $\Gr(\Cc_F)$, then $D[R_{\zeta f}]  = D[S_{\zeta g}]$.\\
(ii) If $[R_f] \cdot [S_g] = [M_m]$ in $\Gr(\Cc_F)$ then $D[R_{\zeta f}] D[S_{\zeta g}]  = D[M_{\zeta m}]$.\\
\end{Cor}

\begin{Rem}\label{rem:pert-def}
(i)  If all $R_f \in \Cc_F$ have finite integrals, then Corollary \ref{cor:rep-Gr} says that the map $[R_f] \mapsto D[R_{\zeta f}]$ defines a ring homomorphism $\Gr(\Cc_F) \rightarrow \End(\Hc)[\![\zeta]\!]$. Since $D[R_{\zeta f}]$ commutes with $L_0$ and $\Lb_0$ (and in fact with all modes of the anti-holomorphic copy of the chiral algebra) the `representation' of $\Gr(\Cc_F)$ on $\Hc$ splits into an infinite direct sum of subrepresentations. One may then wonder why one should consider all of them together, rather than restricting one's attention to a given eigenspace. One reason to do this is that one expects $D[R_f]$ to have the following appealing behaviour under modular transformations. Let $Z[R_f](\tau) = \tr_{\Hc} q{}^{L_0-c/24} (q^*){}^{\Lb_0-c/24} D[R_f]$, where $q = \exp(2 \pi i \tau)$, and let us assume that the infinite sum in $D[R_f]$ converges, and that the trace over $\Hc$ converges for $\tau$ in the upper half plane. The resulting power series in $q$ and $q^*$ will typically not have integral coefficients. But when expressed in terms of $\tilde q = \exp(- 2 \pi i / \tau)$ and $\tilde q{}^*$ we are counting the states that live on a circle intersected by the perturbed defect, and so we expect that
\be
  Z[R_f](\tau) = \sum_{(x,y) \in \Cb \times \Cb} n[R_f]_{x,y} \cdot \tilde q{}^{\,x} (\tilde q^*){}^{y}
  \quad ,~n[R_f]_{x,y} \in \Zb_{\ge 0} ~,
\ee
and $n[R_f]_{x,y} \neq 0$ only for countably many pairs. The infinite direct sum of subrepresentations on $\Hc$ has to conspire in a precise way in order to give rise to non-negative integer coefficients in the crossed channel.
\\[.3em]
(ii) The construction of perturbed topological defects and their relation to $\Cc_F$ applies also to perturbations of conformal boundary conditions. Of course, in this case the composition in Theorem \ref{thm:D-prop}\,(ii) does not make sense, but Theorem \ref{thm:D-prop}\,(i) remains valid. In the Cardy case, the discussion of perturbed boundary conditions is however subsumed in that of perturbed topological defects because (in the Cardy case) the boundary state of a perturbed boundary condition can always be written as $D[R_f]| \one \rangle\!\rangle$ for $| \one \rangle\!\rangle$ the Cardy boundary state \cite{Cardy:1989ir} associated to the vacuum representation of $V$. This follows from the 3d\,TFT formulation of boundary and defect correlators \cite{Felder:1999mq,tft4}. So in the Cardy case, treating perturbed conformal boundaries instead of perturbed topological defects amounts to forgetting the monoidal structure on $\Cc_F$.
\end{Rem}

\section{Example: Lee-Yang model}\label{sec:LY}

\subsection{Bulk theory and perturbed defects}\label{sec:LY-bulk}

The Lee-Yang model is the Virasoro minimal model $M(2,5)$ of central charge $c=-22/5$. The two irreducible highest weight representations of the Virasoro algebra that lie in the Kac table have highest weights $h_{(1,1)} = h_{(1,4)} = 0$ and $h_{(1,2)} = h_{(1,3)} = -1/5$. We will abbreviate $1=(1,1)$ and $\phi = (1,2)$, and we will denote the corresponding representations by $R_1$ (for $h=0$) and $R_\phi$ (for $h=-1/5$).
As already remarked in footnote \ref{fn:Ri-vs-Rf-notation}, the notation $R_1$ and $R_\phi$ should not be confused with objects $R_f$ of $\mathcal{C}_F$ (for some $\mathcal{C}$ and $F$); in any case we will never use $1$ or $\phi$ to denote morphisms.

Let $\Rep(V_{2,5})$ be the category of all Virasoro representations at $c=-22/5$ which are isomorphic to finite direct sums of $R_1$ and $R_\phi$. On $\Rep(V_{2,5})$ we have the fusion tensor product\footnote{
  More precisely, $V_{2,5}$ is the Virasoro vertex operator algebra built on $R_1$. $\Rep(V_{2,5})$ is the category of admissible modules of $V_{2,5}$; this category is finite and semi-simple \cite[Def.\,2.3\,\&\,Thm.\,4.2]{Wang:1993} and forms a braided monoidal category \cite[Cor.\,3.9]{Huang:1995}.}
with non-trivial fusion $R_\phi \otimes R_\phi \cong R_1 \oplus R_\phi$. The Grothendieck group of $\Rep(V_{2,5})$ is therefore isomorphic to $\Zb \times \Zb$ with generators $[R_1]$ and $[R_\phi]$. The product on $\Gr(\Rep(V_{2,5}))$ has $[R_1]$ as multiplicative unit, and $[R_\phi] \cdot [R_\phi] = [R_1] + [R_\phi]$.

\medskip

The characters of $R_1$ and $R_\phi$ are (see e.g.\ \cite{Nahm:2004ch})
\be
 \begin{array}{lll} \displaystyle
  \chi_1(\tau) = \tr_{R_1} q^{L_0-c/24} \etb= q^{11/60} \hspace{-1.2em}\prod_{n \equiv 2,3 \,\text{mod}\, 5}\hspace{-1.1em} (1-q^n)^{-1} \mem
  \etb= q^{11/60}( 1+q^2+q^3+q^4 + \dots)~,
  \\[1.5em] \displaystyle
  \chi_\phi(\tau) = \tr_{R_\phi} q^{L_0-c/24} \etb= q^{-1/60} \hspace{-1.2em}\prod_{n \equiv 1,4 \,\text{mod}\, 5}\hspace{-1.1em} (1-q^n)^{-1} \mem
  \etb= q^{-1/60}( 1+q+q^2+q^3+2q^4 + \dots)~,
\end{array}
\ee
where $q = e^{2 \pi i \tau}$ and the products are from $n=1$ to infinity with the restriction mod 5 as shown. Under the modular transformation $\tau \mapsto -1/\tau$ they transform as $\chi_a(-1/\tau) = \sum_{b \in \{1,\phi\}} S_{ab} \, \chi_b(\tau)$ with
\be
  S = \begin{pmatrix} S_{11} & S_{1\phi} \\ S_{\phi1} & S_{\phi\phi} \end{pmatrix}
  = \frac{-1}{|\sqrt{d{+}2}\,|}\begin{pmatrix} 1 & d \\ d & -1 \end{pmatrix}~, \quad
   \text{where } d = \frac{1-\sqrt{5}}2 = -0.618... ~.
\labl{eq:LY-Smat}
The space of states of the Lee-Yang model is
\be
  \Hc = R_1 \otimes_\Cb R_1 \, \oplus \, R_\phi \otimes_\Cb R_\phi ~.
\ee
The partition function $Z(\tau) = \tr_\Hc( q{}^{L_0-c/24} (q^*){}^{\Lb_0 -c/24}) = |\chi_1(\tau)|^2 + |\chi_\phi(\tau)|^2$ is modular invariant, as it should be.

As described in Section \ref{sec:top-def}, to each object in $R \in \Rep(V_{2,5})$ we can associate a topological defect operator $D[R] : \Hc \rightarrow \Hc$ that commutes with the two copies of the Virasoro algebra. Since $D[R]$ depends only on $[R] \in \Gr(\Rep(V_{2,5}))$, it is enough to give $D[R_1]$ and $D[R_\phi]$ as in \eqref{eq:top-def-op},
\be
  D[R_1] = \id_\Hc ~~,~~~
  D[R_\phi] = d \cdot \id_{R_1 \otimes_\Cb R_1} -\, d^{-1} \cdot \id_{R_\phi \otimes_\Cb R_\phi} ~,
\ee
where $d$ is as in \eqref{eq:LY-Smat}. It is easy to check that indeed $D[R_\phi]D[R_\phi] = \id + D[R_\phi]$, as required by the corresponding relation in $\Gr(\Rep(V_{2,5}))$.

We can now perturb the defect labelled $R_\phi$ by a chiral defect field with left/right conformal weights $(-\tfrac15,0)$ as described in Section \ref{sec:pert-def}. This amounts to considering the objects $R_\phi(\mu)\equiv (R_\phi, \mu\cdot\lambda_{(\phi\phi)\phi})$ in $\Cc_{R_\phi}$, where $\mu\in\mathbb{C}$ and $\lambda_{(\phi\phi)\phi}$ is a fixed non-zero morphism $R_\phi\otimes R_\phi\rightarrow R_\phi$. We then obtain a family of defect operators $D[R_\phi(\lambda)]$. In \cite{Runkel:2007wd} it was shown -- assuming convergence -- that these operators mutually commute,
\be
  \big[\,D[R_\phi(\lambda)]\,,\,D[R_\phi(\mu)]\,\big] = 0
  \qquad \text{for all}~\lambda,\mu\in\Cb~,
\ee
and that they satisfy the functional relation
\be
  D[R_\phi(e^{2\pi i / 5} \lambda)] \, D[R_\phi(e^{-2\pi i / 5} \lambda)] = \id + D[R_\phi(\lambda)]
  \qquad \text{for all}~\lambda\in\Cb~.
\labl{eq:LY-fun-rel}
In the next section we recover this functional relation from studying the tensor product and exact sequences in the corresponding category $\Cc_F$.

\subsection[The category $\Cc_F$ for the Lee-Yang model]{The category $\boldsymbol{\Cc_F}$ for the Lee-Yang model}\label{sec:LY-ex}

The category $\Rep(V_{2,5})$ is equivalent (as a $\Cb$-linear braided monoidal category) to a category $\Vc$ defined as follows. The objects $A$ of $\Vc$ are pairs $A = (A_1,A_\phi)$ of finite-dimensional complex vector spaces indexed by the labels $\{1,\phi\}$ used for simple objects in $\Rep(V_{2,5})$. A morphism $f : A \rightarrow B$ is a pair $f = (f_1,f_\phi)$ of linear maps, where $f_1 : A_1 \rightarrow B_1$ and $f_\phi : A_\phi \rightarrow B_\phi$. This construction is described in more detail in Appendix \ref{app:catV}.
The tensor product $\circledast$ of $\Vc$ is given on objects as
\be
  A \circledast B = \big(~
  A_1 \otimes_\Cb B_1 \,\oplus\,  A_\phi \otimes_\Cb B_\phi ~,~
  A_1 \otimes_\Cb B_\phi \,\oplus\,  A_\phi \otimes_\Cb B_1 \,\oplus\,  A_\phi \otimes_\Cb B_\phi ~\big) ~.
\labl{eq:LY-AB-tens}
The tensor product on morphisms and the non-trivial associator are described in Appendix \ref{app:catV}. The dual of an object $A \in \Vc$ is $A^\vee = (A_1^*, A_\phi^*)$, where $A_1^*$ and $A_\phi^*$ are the dual vector spaces. The duality morphisms are given in Appendix \ref{app:catV}.

As representatives of the two isomorphism classes of simple objects we take $\one = (\Cb,0)$ and $\Phi = (0,\Cb)$. We are interested in the category $\Vc_F$ for $F= \Phi$. Note that $\Phi \circledast A = ( A_\phi , A_1 \oplus A_\phi )$. Therefore, in an object $A_f \in \Vc_\Phi$, the morphism $f : \Phi \circledast A \rightarrow A$ has components $f_1 : A_\phi \rightarrow A_1$ and $f_\phi : A_1 \oplus A_\phi \rightarrow A_\phi$.
We will denote the two summands of $f_\phi$ as $f_{\phi 1} : A_1 \rightarrow A_\phi$ and $f_{\phi \phi} : A_\phi \rightarrow A_\phi$; for consistency of notation we will also denote $f_1 \equiv f_{1\phi}$. It is convenient to collect these three linear maps into a matrix
\be
  f = \raisebox{.8em}{$
  \begin{array}{cc}
    & \!\!\!\! \begin{array}{cc} A_1 & A_\phi \end{array} \\
    \begin{array}{c} A_1 \\ A_\phi \end{array} &
    \!\!\!\! \begin{pmatrix} 0 & f_{1 \phi} \\ f_{\phi 1} & f_{\phi\phi} \end{pmatrix}
  \end{array}  $}
  ~~,
\labl{eq:LY-f-PhiA-A}
where we have also indicated the source and target vector spaces. We can now compute the dual of an object $A_f \in \Vc_\Phi$ according to \eqref{eq:cf-def}. This is done in Appendix \ref{app:T-c-LY} with the simple result
\be
  (A_f)^\vee = (A^\vee, c(f))
  \quad \text{with} \quad
  c(f) = \raisebox{.8em}{$
  \begin{array}{cc}
    & \!\!\!\! \begin{array}{cc} A_1^*~~~~~ & ~~~~~A_\phi^* \end{array} \\[.3em]
    \begin{array}{c} A_1^* \\ A_\phi^* \end{array} &
    \!\!\!\! \begin{pmatrix} 0 & - d \zeta^2 \, f_{\phi 1}^* \\
    - d^{-1} f_{1 \phi}^* & - \zeta \, f_{\phi\phi}^* \end{pmatrix}
  \end{array}  $}
    ~~ \text{and} ~~ \zeta=e^{-\pi i/5}  ~~.
\labl{eq:LY-dual-c}
The tensor product in $\Vc_\Phi$ is more lengthy. We have $A_f \hat\circledast B_g = (A \circledast B, T(f,g))$ where $T(f,g) : \Phi \circledast (A \circledast B) \rightarrow A \circledast B$. The source vector spaces of $T(f,g)$ are (we omit the `$\otimes_\Cb$')
\be
  \Phi \circledast (A \circledast B) = (
  A_1  B_\phi \oplus A_\phi  B_1 \oplus A_\phi  B_\phi ,
  A_1  B_1 \oplus A_\phi  B_\phi \oplus A_1  B_\phi \oplus A_\phi  B_1 \oplus A_\phi  B_\phi )~.
\ee
In Appendix \ref{app:T-c-LY} we evaluate equation \eqref{eq:T-def} for $T(f,g)$ in the category $\Vc_\Phi$. The result is best represented in a $5 \times 5$-matrix, again omitting `$\otimes_\Cb$',
\be
  \begin{array}{cc}
    & \!\!\!\!
    \begin{array}{ccccc}
    A_1  B_1 ~~~~ & ~~~
    A_\phi  B_\phi ~~~ & ~~~
    A_1  B_\phi ~~~ & ~~~
    A_\phi  B_1 ~~~~~~ & ~~
    A_\phi  B_\phi ~~~
    \end{array} \\[.8em]
    \begin{array}{c}
    A_1  B_1
    \\[.8em]
    A_\phi  B_\phi
    \\[.8em]
    A_1  B_\phi
    \\[.8em]
    A_\phi  B_1
    \\[.8em]
    A_\phi  B_\phi
    \end{array}
    &
    \!\!\!\! \begin{pmatrix}
    0 & 0 & \id_{A_1} g_{1\phi} & f_{1\phi} \id_{B_1} & 0
    \\[.8em]
    0 & 0 & f_{\phi 1} \id_{B_\phi} & \zeta^2 \id_{A_\phi} g_{\phi 1} & f_{\phi\phi} + \zeta g_{\phi\phi}
    \\[.8em]
    \id_{A_1} g_{\phi 1} & \tfrac{1}{d} f_{1\phi}\id_{B_\phi} & \id_{A_1} g_{\phi\phi} & 0 & w  f_{\phi 1} \id_{B_\phi}
    \\[.8em]
    f_{\phi1} \id_{B_1} & \tfrac{1}{\zeta^2 d} \id_{A_\phi} g_{1 \phi} & 0 & f_{\phi\phi} \id_{B_1} & \tfrac{w}{\zeta} \id_{A_\phi} g_{1\phi}
    \\[.8em]
    0 & \tfrac{1}{wd}(f_{\phi\phi}{+}\tfrac{1}{\zeta} g_{\phi\phi}) & f_{\phi 1} \id_{B_\phi} & \zeta \id_{A_\phi} g_{\phi 1} & -\tfrac{1}{d}(f_{\phi\phi}{+}g_{\phi\phi})
    \end{pmatrix}
  \end{array}  ~~.
\labl{eq:LY-Tfg}
Here $\zeta$ was given in \eqref{eq:LY-dual-c}, $w \in  \Cb^\times$ is a normalisation constant (see Appendix \ref{app:T-c-LY}), and in the entries with sums we have omitted the identity maps. For example, $f_{\phi\phi} + \zeta g_{\phi\phi}$ stands for $f_{\phi\phi} \otimes_\Cb \id_{B_\phi} + \zeta \id_{A_\phi} \otimes_\Cb g_{\phi\phi}$.

\subsection[Some exact sequences in $\Cc_F$]{Some exact sequences in $\boldsymbol{\Cc_F}$}\label{sec:LY-ex-seq}

Two objects $A_f$ and $B_g$ in $\Vc_\Phi$ are isomorphic if and only if there exist isomorphisms $\gamma_1 : A_1 \isorightarrow B_1$ and $\gamma_\phi : A_\phi \isorightarrow B_\phi$ such that
\be
  \begin{pmatrix} 0 & g_{1\phi} \\ g_{\phi 1} & g_{\phi\phi} \end{pmatrix}
  =  \begin{pmatrix} 0 & \gamma_1 \circ f_{1\phi} \circ \gamma_\phi^{-1} \\ \gamma_\phi\circ  f_{\phi 1} \circ \gamma_1^{-1} & \gamma_\phi\circ  f_{\phi\phi} \circ \gamma_\phi^{-1} \end{pmatrix} ~.
\ee
For $\lambda \in \Cb$ write $\Phi(\lambda) \equiv (\Phi, f(\lambda))$ with $f(\lambda)_1=0$ and $f(\lambda)_\phi = \lambda \cdot \id_\Cb$. In other words, $\Phi(\lambda) = \big( (0,\Cb) , (\lambda) \big)$. Then $\Phi(\lambda) \cong \Phi(\mu)$ if and only if $\lambda=\mu$. As another example,
\be
  \Bigg(\! (\Cb, \Cb) , \begin{pmatrix} 0 & a \\ b & c \end{pmatrix} \mem\Bigg)
  \cong
  \Bigg(\! (\Cb, \Cb) , \begin{pmatrix} 0 & a' \\ b' & c' \end{pmatrix} \mem\Bigg)
  ~\Leftrightarrow~
  \begin{cases}
  ab=a'b' ~,~ c = c'  ~\text{and} \\
  \mathrm{rk}(a) =
  \mathrm{rk}(a') ~,~
  \mathrm{rk}(b) =
  \mathrm{rk}(b')
  \end{cases} ~,
\labl{eq:11-iso}
where $\mathrm{rk}(a) \in \{0,1\}$ denotes the rank of the linear map $a \cdot \id_\Cb$.

\medskip

For $\one$ and $\Phi(\lambda)$ there are no non-trivial exact sequences as the underlying objects in $\Vc$ are already simple. For $\big( (\Cb, \Cb) , \big(\begin{smallmatrix} 0 & a \\ b & c \end{smallmatrix}\big) \big)$ there are two exact sequences,
\be
  0 \rightarrow \Phi(\lambda) \rightarrow \big( (\Cb, \Cb) , \big(\begin{smallmatrix} 0 & 0 \\ b & \lambda \end{smallmatrix}\big) \big) \rightarrow \one \rightarrow 0
  ~~,~~
  0 \rightarrow \one \rightarrow \big( (\Cb, \Cb) , \big(\begin{smallmatrix} 0 & a \\ 0 & \lambda \end{smallmatrix}\big) \big) \rightarrow \Phi(\lambda) \rightarrow 0 ~.
\labl{eq:LY-C-C-seq}
Let us explain how one arrives at the first one. One checks that there is a surjective morphism $\big( (\Cb, \Cb) , \big(\begin{smallmatrix} 0 & a \\ b & c \end{smallmatrix}\big) \big) \rightarrow \one$ in $\Vc_\Phi$ iff $(1,0) \big(\begin{smallmatrix} 0 & a \\ b & c \end{smallmatrix}\big) = 0$, i.e.\ iff $a=0$. To complete this to an exact sequence, we need an injective morphism $\Phi(\lambda) \rightarrow\big( (\Cb, \Cb) , \big(\begin{smallmatrix} 0 & a \\ b & c \end{smallmatrix}\big) \big)$. This exists iff $\big(\begin{smallmatrix} 0 & a \\ b & c \end{smallmatrix}\big)\big(\begin{smallmatrix} 0 \\ 1 \end{smallmatrix}\big) = \big(\begin{smallmatrix} 0 \\ \lambda \end{smallmatrix}\big)$, i.e.\ iff $a=0$ and $\lambda=c$. From \eqref{eq:LY-C-C-seq} it follows that in $\Gr(\Vc_\Phi)$ we have
\be
 \big[\,\big( (\Cb, \Cb) ,  \big(\begin{smallmatrix} 0 & 0 \\ b & \lambda \end{smallmatrix}\big) \big) \,\big]
  ~=~ [ \one ] + [\Phi(\lambda)] ~=~
  \big[\, \big( (\Cb, \Cb) , \big(\begin{smallmatrix} 0 & a \\ 0 & \lambda \end{smallmatrix}\big) \big) \,\big] ~,
\ee
even though $\big( (\Cb, \Cb) ,  \big(\begin{smallmatrix} 0 & 0 \\ b & \lambda \end{smallmatrix}\big) \big)$ and $\big( (\Cb, \Cb) , \big(\begin{smallmatrix} 0 & a \\ 0 & \lambda \end{smallmatrix}\big) \big)$ are not isomorphic unless $a=b=0$.

\medskip

Next let us look at the simplest non-trivial tensor product, $\Phi(\lambda) \hat\circledast \Phi(\mu)$.  Formula \eqref{eq:LY-Tfg} simplifies to
\be
  \Phi(\lambda) \hat\circledast \Phi(\mu) =   \Bigg(\! (\Cb, \Cb) , \begin{pmatrix} 0 & \lambda + \zeta \mu \\ \tfrac{1}{wd}(\lambda+\zeta^{-1}\mu) & -d^{-1}(\lambda+\mu) \end{pmatrix} \mem\Bigg) ~.
\labl{eq:PP-tens-LY}
By comparing to \eqref{eq:11-iso} we see that $\Phi(\lambda) \hat\circledast \Phi(\mu) \cong \Phi(\mu) \hat\circledast \Phi(\lambda)$ iff either $\lambda=\mu=0$ or $(\lambda + \zeta \mu)(\lambda + \zeta^{-1} \mu) \neq 0$. In particular, $\Phi(-\zeta \mu) \hat\circledast \Phi(\mu) \ncong \Phi(\mu) \hat\circledast \Phi(-\zeta \mu)$ unless $\mu=0$. This shows that $\Vc_\Phi$ cannot be braided. The reducibility of $\Phi(\lambda) \hat\circledast \Phi(\mu)$ is summarised in three cases:
\\[.3em]
(i) if $\lambda \notin \{-\zeta \mu, -\zeta^{-1}\mu\}$ then $\Phi(\lambda) \hat\circledast \Phi(\mu)$ is irreducible,
\\[.3em]
(ii) if $\lambda = - \zeta \mu$ we have $0 \rightarrow \Phi(\zeta^{-2} \mu) \rightarrow \Phi(-\zeta \mu) \hat\circledast \Phi(\mu) \rightarrow \one \rightarrow 0$,
\\[.3em]
(iii) if $\lambda = - \zeta^{-1} \mu$ we have $0 \rightarrow \one \rightarrow \Phi(-\zeta^{-1} \mu) \hat\circledast \Phi(\mu) \rightarrow \Phi(\zeta^2 \mu) \rightarrow 0$.
\\[.3em]
In $\Gr(\Vc_\Phi)$ we therefore get the relations
\be
  [\Phi(\zeta^{-2} \lambda)] \cdot [\Phi(\zeta^{2}\lambda)]
  ~\overset{\text{(ii)}}=~ [1] + [\Phi(\lambda)] ~\overset{\text{(iii)}}=~
  [\Phi(\zeta^2 \lambda)] \cdot [\Phi(\zeta^{-2}\lambda)]  ~.
\labl{eq:PP-rel-Gr}
Combining with the case when $\Phi(\lambda) \hat\circledast \Phi(\mu)$ is irreducible we find that in $\Gr(\Vc_\Phi)$ we have
\be
  [\Phi(\lambda)] \cdot [\Phi(\mu)]  =  [\Phi(\mu)] \cdot [\Phi(\lambda)]
  \quad \text{for all} ~ \lambda,\mu \in \Cb ~.
\ee

In fact we could have obtained the reducibility in (ii) and (iii) above already from the existence of duals. Namely, by \eqref{eq:LY-dual-c}, $(\Phi(\lambda))^\vee = \Phi(-\zeta\lambda)$ and by Lemma \ref{lem:CF-duals} we have non-zero morphisms $b_\Phi : \one \rightarrow \Phi(\lambda) \Phi(- \zeta \lambda)$ and $d_\Phi : \Phi(-\zeta \lambda) \Phi(\lambda) \rightarrow \one$. Also note that taking the dual $n$-times gives $\Phi(\lambda)^{\vee  \dots \vee} = \Phi( (-\zeta)^n \lambda )$, and since $-\zeta$ is a 10th root of unity, the 10-fold dual is the first one that is again isomorphic to $\Phi(\lambda)$ (for $\lambda \neq 0$). This is different from e.g.\ fusion categories (which are by definition semi-simple \cite[Def.\,1.9]{CalEt-04}) where $V^{\vee\vee} \cong V$ for all simple objects $V$, see \cite[Prop.\,1.17]{CalEt-04}.

To conclude our sample calculations in $\Vc_\Phi$ we point out that for a given $ \big( (\Cb,\Cb),\big(\begin{smallmatrix} 0 & a \\ b & c \end{smallmatrix}\big) \big)$ at least one of the isomorphisms
\be
   \big( (\Cb,\Cb),\big(\begin{smallmatrix} 0 & a \\ b & c \end{smallmatrix}\big) \big) \cong \one \oplus \Phi(\lambda), \
   \big( (\Cb,\Cb),\big(\begin{smallmatrix} 0 & a \\ b & c \end{smallmatrix}\big) \big) \cong \Phi(\lambda) \hat\circledast \Phi(\mu) ~,
\labl{eq:11-DD-prod}
holds for some $\lambda,\mu\in\mathbb{C}$. This is easy to check by comparing cases in \eqref{eq:11-iso} and \eqref{eq:PP-tens-LY}.

\subsection{Some implications for defect flows}

The relation \eqref{eq:PP-rel-Gr} in $\Gr(\Vc_\Phi)$ gives the functional relation \eqref{eq:LY-fun-rel} for the perturbed $R_\phi$-defect in the Lee-Yang model. Let us point out one application of such functional relations, namely how they can give information about endpoints of renormalisation group flows. We use the notation for objects as in $\Vc_\Phi$, e.g.\ we write $D[\Phi(\lambda)]$ instead of $D[R_\phi(\lambda)]$.

We shall assume that $D[\Phi(\lambda)]$ is an operator valued meromorphic function on $\Cb$, and that its asymptotics for $\lambda \rightarrow +\infty$ along the real axis is given by (compare to \cite[Eqn.\,(62)]{Bazhanov:1994ft} or \cite[Eqn.\,(2.21)]{Bazhanov:1996dr})
\be
  D[\Phi(\lambda)] \sim \exp( f \lambda^{1/(1-h_\phi)} ) D_\infty ~+~ \text{less singular terms} ~,
\labl{eq:LY-Dphi-as}
where $\lambda^{1/(1-h_\phi)} = \lambda^{5/6}$ has dimension of length, $f>0$ is a free energy per unit length, and $D_\infty$ is the operator describing the defect at the endpoint of the flow. We assume that this asymptotic behaviour remains valid in the direction $\lambda = r e^{i \theta}$, $r\rightarrow +\infty$, of the complex plane at least as long as the real part of $(e^{i \theta})^{5/6}$ remains positive, i.e.\ for $|\theta|<3 \pi/5$. This is a subtle point as in analogy with integrable models the asymptotics will be subject to Stokes' phenomenon, see e.g.\ \cite[App.~D.1]{Dorey:2007zx}.

With these assumptions, we can substitute the asymptotic behaviour \eqref{eq:LY-Dphi-as} into the functional relation \eqref{eq:LY-fun-rel}, which gives
\be
  \exp\big( f (\zeta^2 \lambda)^{5/6} + f (\zeta^{-2} \lambda)^{5/6} \big) D_\infty D_\infty = \id + \exp( f \lambda^{5/6} ) D_\infty  ~.
\ee
As $f>0$, the identity operator will be subleading, and since $ (\zeta^2)^{5/6} + (\zeta^{-2})^{5/6} = 1$ the leading asymptotics demands that
\be
  D_\infty D_\infty = D_\infty ~.
\labl{eq:Dinf-as}
Since $D_\infty$ is the endpoint of a renormalisation group flow, we expect it to be a conformal defect, i.e.\ $[L_m + \bar L_{-m},D_\infty]=0$. On the other hand for every value of $\lambda$ we have $\big[\bar L_m , D[\Phi(\lambda)]\,\big]=0$, so that $D_\infty$ is again a topological defect. Thus $D_\infty = m \cdot \id + n \cdot D_\phi$ for some $m,n \in \Zb_{\ge 0}$. This is consistent with \eqref{eq:Dinf-as} only for $D_\infty = \id$. We thus obtain the asymptotic behaviour
\be
  D[\Phi(\lambda)] \xrightarrow{~\lambda\rightarrow+\infty~} \exp( f \lambda^{5/6} ) \id ~.
\labl{eq:LY-Dphi-as2}
This is the expected result, because via the relation of perturbed defects and perturbed boundaries mentioned in Remark \ref{rem:pert-def}\,(ii), the above flow agrees with the corresponding boundary flow obtained in \cite[Sect.\,3]{Dorey:1997yg}. It also agrees with the corresponding free field expression \cite[Eqn.\,(2.21)]{Bazhanov:1996dr}.

This result allows us to make some statements about perturbations of the superposition of the $1$- and $\phi$-defect, i.e.\ the topological defect labelled by $R_1\oplus R_\phi$. We can either perturb it by a defect field on the topological defect labelled $R_\phi$ alone, in which case we would get the operator $\id + D[\Phi(\lambda)]$ which flows to $D_\infty = \id$ as $\lambda \rightarrow +\infty$. Or we can in addition perturb by defect changing fields. In this case we can use the result \eqref{eq:11-DD-prod}, which tells us that we can write the perturbed defect as the composition $D[\Phi(\lambda)] D[\Phi(\mu)]$ for some $\lambda,\mu$. Then, if the necessary $\lambda,\mu$ lie in the wedge of the complex plane where \eqref{eq:LY-Dphi-as2} is valid, we again have
\be
  D\Big[  \big( (\Cb,\Cb),\big(\begin{smallmatrix} 0 & ra \\ rb & rc \end{smallmatrix}\big) \big) \Big] \xrightarrow{~r\rightarrow+\infty~} \exp( f' r^{5/6} ) \id  ~.
\ee

\section{Conclusions}\label{sec:C}

In this paper we have proposed an abelian rigid monoidal category $\Cc_F$, constructed from an abelian rigid braided monoidal category $\Cc$ and a choice of object $F \in \Cc$, that captures some of the properties of perturbed topological defects. To make the connection to defects, we set $\Cc = \Rep(V)$, for $V$ a rational vertex operator algebra, and choose a $V$-module $F \in \Cc$ together with a vector $\phi \in F$. Then we consider the charge-conjugation CFT constructed from $V$ (the Cardy case). An object $U_f \in \Cc_F$ corresponds to an unperturbed topological defect labelled $U$ and a perturbing field given by the chiral defect field defined via $\phi \in F$ and the morphism $f: F\otimes U \rightarrow U$. Assuming convergence of the multiple integrals and the infinite sum in \eqref{eq:D[Rf]-def}, to $U_f$ we can assign an operator $D[U_f]$ on the space of states $\Hc = \bigoplus_{i \in \Ic} R_i \otimes_\Cb R_i^\vee$ of the CFT. This operator describes the topological defect perturbed by the specified defect field. Again assuming convergence of all $D[\dots]$ involved, the main properties of the assignment $U_f \mapsto D[U_f]$ are
\begin{itemize} \itemsep 0pt
\item[(i)] $D[\one] = \id_\Hc$,
\item[(ii)] $D[U_{f=0}] = \sum_{i,j \in\Ic} \dim\Hom(R_i,U) \, S_{ij}/S_{0j} \, \id_{R_j \otimes_\Cb R_j^\vee}$,
\item[(iii)] $\big[\,L_0\,,\,D[U_f]\,\big] = 0$ and $\big[\,\Lb_m\,,\,D[U_f]\,\big] = 0$ for $m \in \Zb$,
\item[(iv)] if $0 \rightarrow K_h \rightarrow U_f \rightarrow C_g\rightarrow 0$ is an exact sequence, then $D[U_f] = D[K_h]+D[C_g]$,
\item[(iv${}'$)] if $[U_f] = [V_g]$ in $\Gr(\Cc_F)$, then $D[U_f] = D[V_g]$,
\item[(v)] $D[U_f \hotimes V_g] = D[U_f] D[V_g]$.
\end{itemize}
There is an anti-holomorphic counterpart of the construction in this paper, where one perturbs the topological defect by a defect field of dimension $(0,h)$. This generates another set of defect operators which commute with those introduced here.

\medskip

The results of this paper also leave a large number of question unanswered, and we hope to come back to some of these in the future:
\\[.3em]
{\bf 1.} In the Lee-Yang example it should be possible to describe the category $\Cc_F$ and its Grothen\-dieck ring more explicitly. For example it would be interesting to know if $\Cc_F$ is generated by the $\Phi(\lambda)$ in the sense that every object of $\Cc_F$ is obtained by taking direct sums, tensor products, subobjects and quotients starting from $\Phi(\lambda)$. Note that we do at this stage not even know whether or not $\Cc_F$ is commutative in the Lee-Yang example.
\\[.3em]
{\bf 2.} Consider the case $\Cc = \Rep(V)$ for a rational vertex operator algebra $V$ and let $U_f \in \Cc_F$ have finite integrals. Suppose the infinite sum $O(\zeta) = D[U_{\zeta f}]$ has a finite radius of convergence in $\zeta$. One can then extend the domain of definition of $O(\zeta)$ by analytic continuation. To solve the functional relations it is most important to understand the global properties of $O(\zeta)$, in particular whether all functions $\varphi( O(\zeta) v)$ (for $\varphi \in \Hc^*$ and $v \in \Hc$) are entire functions on $\Cb$, and what their asymptotic behaviours are. It should be possible to address these questions with the methods reviewed and developed in \cite{Dorey:2007zx} and \cite{Inoue:2008}.
\\[.3em]
{\bf 3.} The category $\Cc_F$ is designed specifically for the Cardy case. The formalism developed in $\cite{tft1,defect}$ allows one to extend this treatment to all rational CFTs with chiral symmetry $V \otimes_\Cb V$. The different CFTs with this symmetry are in one-to-one correspondence with Morita-classes of special symmetric Frobenius algebras $A$ in $\Cc = \Rep(V)$. Given such an algebra $A$, the category $\Cc_F$ has to be replaced by a category $\Cc(A)_F$ whose objects are pairs $(B,f)$ where $B$ is an $A$-$A$-bimodule and $f : F \otimes^+ B \rightarrow B$ is an intertwiner of bimodules (see \cite[Sect.\,2.2]{tft4} for the definition of $\otimes^+$). The details remain to be worked out. For $A = \one$ one recovers the Cardy case discussed in this paper.
\\[.3em]
{\bf 4.} It would be interesting to understand if the map $\Gr(\Cc_F) \rightarrow \End(\Hc)$ from the Grothendieck ring to defect operators is injective. The map $\Gr(\Cc) \rightarrow \End(\Hc)$ taking the class $[R]$ of a representation of the rational vertex operator algebra $V$ to the topological defect $D[R]$ is known to be injective, and in fact a corresponding statement holds for symmetry preserving topological defects in all rational CFTs with chiral symmetry $V \otimes_\Cb V$ \cite{Fuchs:2007vk}.
\\[.3em]
{\bf 5.} It would be good to investigate the properties of $\Cc_F$ in more examples. The evident ones are the Virasoro minimal models, the SU(2)-WZW model, the rational free boson, etc. Or, coming from the opposite side, one could use the fact that modular categories with three or less simple objects (and unitary modular categories with four or less simple objects) have been classified \cite{Rowell:2007}, and study $\Cc_F$ for all $\Cc$ in that list and different choices of $F$. The proper treatment of supersymmetry in the present formalism also remains to be worked out.
\\[.3em]
{\bf 6.} One application of the perturbed defect operators is the investigation of boundary flows. As pointed out in Remark \ref{rem:pert-def}\,(ii), in the Cardy case the boundary state of a perturbed conformal boundary condition can be written as $D[U_f] | \one \rangle\!\rangle$. However, for other modular invariants this need not be true. But, as in the unperturbed case \cite[Sect.\,2]{Schweigert:2006af}, the category of perturbed boundary conditions will form a module category over the category of chirally perturbed defect lines. It would be interesting to investigate this situation in cases where the two categories are distinct (as abelian categories).
\\[.3em]
{\bf 7.} In general an object $U_f \in \Cc_F$ describes a topological defect perturbed by defect changing fields. Placed in front of the conformal boundary labelled by the vacuum representation $\one \in \Cc$ one obtains the boundary condition $U$ perturbed by boundary changing fields. Such perturbations have been studied for unitary minimal models in \cite{Graham:2001pp}. While our method is not directly applicable to unitary minimal models (the multiple integrals diverge in this case as $h_{1,3}\ge\tfrac12$), one could still study if the functional relations predict a similar flow pattern for the non-unitary models.
\\[.3em]
{\bf 8.} The relation to finite-dimensional representations of quantum affine algebras should be worked out beyond the remarks in Appendix \ref{sec:CF-Uq}.
\\[.3em]
{\bf 9.} Baxter's $Q$-operator is a crucial tool in the solution of integrable lattice models. Such $Q$-operators have been obtained in chiral conformal field theory \cite{Fendley:1995,Bazhanov:1996dr,Bazhanov:1998dq}, and in lattice models via the representation theory of quantum affine algebras \cite{Kuniba:1993cn,Rossi:2002,Korff:2003}. Recently they have also been studied in certain (discretised) non-rational conformal and massive field theories \cite{Bytsko:2009mg}. It would be good to translate these constructions and obtain $Q$-operators also in the present language.

\newpage

\appendix

\section{Appendix}

\subsection{Relation to evaluation representations of quantum affine sl(2)}\label{sec:CF-Uq}

In this appendix we collect some preliminary remarks on the relation of a category of the form $\Cc_F$ and evaluation representations of the quantum affine algebra $U_q(\widehat{sl}(2))$. We follow the conventions of \cite{Chari:1991}. Let $q \in \Cb^\times$ be not a root of unity. The quantum group $U_q(sl(2))$ is generated by elements $e^{\pm}$, $K^{\pm 1}$ with relations
\be
  K K^{-1} = K^{-1} K = 1
  ~~,~~~
  K e^{\pm} K^{-1} = q^{\pm 2} e^\pm
  ~~,~~~
  [e^+,e^-] = \frac{K-K^{-1}}{q-q^{-1}} ~.
\ee
The quantum group $U_q(\widehat{sl}(2))$ is generated by elements $e^{\pm}_i$, $K^{\pm 1}_i$, $i = 0,1$, with relations
\be
  K_i K^{-1}_i = K^{-1}_i K_i = 1
  ~~,~~~
  K_i e^{\pm}_i K^{-1}_i = q^{\pm 2} e^\pm_i
  ~~,~~~
  [e^+_i,e^-_i] = \frac{K_i-K^{-1}_i}{q-q^{-1}}
  ~,
\ee
as well as, for $i \neq j$,
\bea
  [K_0,K_1] = 0
  ~~,~~~
  [e^\pm_0,e^{\mp}_1]=0
  ~~,~~~
  K_i e^\pm_j K^{-1}_i = q^{\mp 2} e^\pm_j
\enl
  (e^\pm_i)^3 e_j^\pm  - e^\pm_j(e_i^\pm)^3 = \frac{q^3-q^{-3}}{q-q^{-1}} \Big(
  (e^\pm_i)^2 e^\pm_j e^\pm_i - e^\pm_i e^\pm_j (e^\pm_i)^2 \Big)  ~.
\eear\ee
Let us abbreviate $U \equiv U_q(sl(2))$ and $\hat U \equiv U_q(\widehat{sl}(2))$.
There are infinitely many ways in which $U$ is a subalgebra of $\hat U$. We will make use of the injective algebra homomorphism $\iota_1 : U \hookrightarrow \hat U$ given by (this is the case $i=0$ in \cite[Sect.\,2.4]{Chari:1991})
\be
  \iota_1(K^{\pm 1}) = K_1^{\pm 1} ~~,~~~
  \iota_1(e^\pm) = e_1^\pm ~.
\ee
This turns $\hat U$ into an infinite-dimensional representation of $U$. Let $\Cc$ be the category of (not necessarily finite-dimensional) representations of $U$. The coproduct of $U$ gives rise to a tensor product on $\Cc$ and the $R$-matrix of $U$ to a braiding.

For each $a \in \Cb^\times$, there is a surjective algebra homomorphism $\mathrm{ev}_a : \hat U \rightarrow U$, described in \cite[Sect.\,4]{Chari:1991}. It has the property that $\mathrm{ev}_a \circ \iota_1 = \id_U$. An {\em evaluation representation} of $\hat U$ is a pull-back of a representation $V$ of $U$ via $\mathrm{ev}_a$ for some $a \in \Cb^\times$. We denote this representation of $\hat U$ by $V(a)$. Let $\mathcal{D}$ be the category of (not-necessarily finite-dimensional) evaluation representations of $\hat U$.

\begin{Thm}
$\mathcal{D}$ is a full subcategory of $\Cc_{\hat U}$.
\end{Thm}

\begin{proof}
Define a map $G$ from $\mathcal{D}$ to $\Cc_{\hat U}$ on objects by $G(V(a)) = (V, \mathrm{ev}_a \otimes_U \id_V)$, where we identified $U \otimes_U V \equiv V$.
We will show that $f : V(a) \rightarrow W(b)$ is a morphism in $\mathcal{D}$ iff $f$ is a morphism $G(V(a)) \rightarrow G(W(b))$ in $\Cc_{\hat U}$. Indeed, the condition for $f$ to be an intertwiner  $f : V(a) \rightarrow W(b)$ is that for all $u \in \hat U$ and $v \in V$ we have
\be
  \mathrm{ev}_b(u).f(v) = f(\mathrm{ev}_a(u).v) ~,
\labl{eq:Uq-morph-check}
and the condition for $f$ to be a morphism $(V,\mathrm{ev}_a \otimes_U \id_V) \rightarrow (W,\mathrm{ev}_b \otimes_U \id_W)$ is
\be
  (\mathrm{ev}_b \otimes_U \id_W) \circ (\id_{\hat U} \otimes_U f)
  = f \circ (\mathrm{ev}_a \otimes_U \id_V) ~.
\ee
If we evaluate this equality on $u \otimes_U v$ for $u \in \hat U$, $v \in V$, we obtain exactly \eqref{eq:Uq-morph-check}.
Thus we can define $G$ on morphisms as $G(f)=f$. It is clear that $G$ is compatible with composition, and that it is full.
\end{proof}
Since $\Cc$ is abelian braided monoidal with exact tensor product, $\Cc_{\hat U}$ is abelian and monoidal by Theorem \ref{thm:CF-tens}.
Let $(\Cc_{\hat U})_f$ be the full subcategory of $\Cc_{\hat U}$ formed by all $(V,g)$ where $V$ is a finite-dimensional representation of $U$. Note that $(\Cc_{\hat U})_f$ is again an abelian
monoidal category. Let $\Rep_f(\hat U)$ be the abelian monoidal category of all finite-dimensional representations of $\hat U$ of type (1,1) (as defined in \cite[Sect.\,3.2]{Chari:1991}). It would be interesting to understand the precise relation between $(\Cc_{\hat U})_f$ and $\Rep_f(\hat U)$. For example, one might expect that $\Rep_f(\hat U)$ is a full subcategory of $(\Cc_{\hat U})_f$.

As a first step towards this goal, one could use that all finite-dimensional irreducible representations of $\hat U$ of type (1,1) are isomorphic to tensor products of evaluation representations \cite[Sect.\,4.11]{Chari:1991}. However, to make use of this property one first has to establish that the tensor product of $\hat U$-representations is compatible with $\hotimes$ defined on $(\Cc_{\hat U})_f$ via the tensor product and braiding on $\Cc$. We do not attempt this in the present paper but hope to return to this point in future work.

\subsection{Proof of Theorem \ref{thm:ab} and Lemma \ref{lem:CF-exact}}\label{app:proof-ab}

In this appendix, $\Cc$ satisfies the assumptions of Theorem \ref{thm:ab}. Namely, $\Cc$ is an abelian monoidal category with right-exact tensor product.

\begin{Lem}\label{lem:CCF-ker-cok}
Let $x : U_f \rightarrow V_g$ and $y : V_g \rightarrow W_h$ be morphisms in $\Cc_F$.\\
(i) If $x : U \rightarrow V$ is a kernel of $y$ in $\Cc$, then $x : U_f \rightarrow V_g$ is a kernel of $y$ in $\Cc_F$.\\
(ii) If $y : V \rightarrow W$ is a cokernel of $x$ in $\Cc$, then $y : V_g \rightarrow W_h$ is a cokernel of $x$ in $\Cc_F$.
\end{Lem}

\begin{proof}
(i) We need to show that $x$ has the universal property of $\ker y$ in $\Cc_F$, namely that for every $U'_{f'} \in \Cc_F$ and every $c : U'_{f'} \rightarrow V_g$ such that $y \circ c = 0$ there exists a unique $\tilde c : U'_{f'} \rightarrow U_f$ such that $c = x \circ \tilde c$. Since $x = \ker y$ in $\Cc$ we know that there exists a unique $\tilde c : U' \rightarrow U$ such that $c = x \circ \tilde c$. It remains to prove that $\tilde c$ is a morphism in $\Cc_F$, i.e.\ that $\tilde{c}\circ f'=f\circ (\id_F\otimes \tilde c)$. To this end consider the following diagram in $\Cc$:
\be
\begin{xy}
(0,30)*+{F\otimes U}="fu"; (30,30)*+{F\otimes V}="fv"; (60,30)*+{F\otimes W}="fw";%
(0,15)*+{U}="u"; (30,15)*+{V}="v"; (60,15)*+{W}="w";%
(30,0)*+{U'}="u'";%
(30,-15)*+{F\otimes U'}="fu'";%
{\ar "fu";"u"};?*!/_.6em/{f};{\ar "fv";"v"};?*!/^2mm/{g};{\ar "fw";"w"};?*!/_.6em/{h};%
{\ar "fu";"fv"}?*!/_.6em/{\id_F\otimes x};{\ar "fv";"fw"}?*!/_.6em/{\id_F\otimes y};
{\ar "u";"v"}?*!/_.6em/{x};{\ar "v";"w"}?*!/_.6em/{y};
{\ar "u'";"v"};?*!/_.6em/{c};%
{\ar "fu'";"u'"};?*!/_.6em/{f'};%
{\ar@{-->} "u'";"u"};?*!/_.6em/{\tilde{c}};
{\ar@{->}@/_2pc/_{\id_F\otimes c} "fu'";"fv"};%
{\ar@{-->}@/^4pc/^{\id_F\otimes \tilde{c}} "fu'";"fu"};
\end{xy}
\ee
By assumption the two triangles commute, and all squares but the one with the two dashed arrows. To establish that also the latter commutes, since $x$ is monic it is enough to show that $x\circ\tilde c \circ f'=x\circ f\circ (\id_F\otimes \tilde c)$. Indeed,
\be
  x\circ f\circ (\id_F\otimes \tilde c)
  = g \circ (\id_F \otimes x) \circ (\id_F\otimes \tilde c)
  = g \circ (\id_F\otimes c)
  = c \circ f'
  = x\circ\tilde{c}\circ f' ~.
\ee
(ii) The proof works along the same lines as that of part (i), but, as opposed to part (i) here we need to use that the tensor product of $\Cc$ is right-exact. For this reason we spell out the details once more.
We need to show that $y$ has the universal property of $\cok x$ in $\Cc_F$. Given a $W'_{h'}$ and a morphism $c : V_g \rightarrow W'_{h'}$ such that $c\circ x=0$, since $y = \cok x$ in $\Cc$ we know there exists a unique morphism $\tilde c : W \rightarrow W'$ in $\Cc$ such that $c = \tilde c \circ y$. It remains to show that $\tilde c : W_h \rightarrow W'_{h'}$ is a morphism in $\Cc_F$, i.e.\ that $\tilde c \circ h = h' \circ (\id_F \otimes \tilde c)$. Consider the diagram:
\be
\begin{xy}
(0,30)*+{F\otimes U}="fu"; (30,30)*+{F\otimes V}="fv"; (60,30)*+{F\otimes W}="fw";%
(0,15)*+{U}="u"; (30,15)*+{V}="v"; (60,15)*+{W}="w";%
(30,0)*+{W'}="w'";%
(30,-15)*+{F\otimes W'}="fw'";%
{\ar "fu";"u"};?*!/_.6em/{f};{\ar "fv";"v"};?*!/_.6em/{g};{\ar "fw";"w"};?*!/^2mm/{h};%
{\ar "fu";"fv"}?*!/_.6em/{\id_F\otimes x};{\ar "fv";"fw"}?*!/_.6em/{\id_F\otimes y};
{\ar "u";"v"}?*!/_.6em/{x};{\ar "v";"w"}?*!/_.6em/{y};
{\ar "v";"w'"};?*!/_.6em/{c};%
{\ar "fw'";"w'"};?*!/^.6em/{h'};%
{\ar@{-->} "w";"w'"};?*!/_.6em/{\tilde{c}};
{\ar@{->}@/_2pc/_{\id_F\otimes c} "fv";"fw'"};%
{\ar@{-->}@/^4.5pc/^{\id_F\otimes \tilde{c}} "fw";"fw'"};
\end{xy}
\ee
Since $y$ is an epimorphism and the tensor product is right-exact, also $\id_F \otimes y$ is an epimorphism. It is therefore enough to show that
$\tilde c \circ h \circ (\id_F \otimes y) = h' \circ (\id_F \otimes \tilde c) \circ (\id_F \otimes y)$. Indeed,
\be
h' \circ (\id_F \otimes \tilde{c})\circ(\id_F\otimes y)
=h'\circ (\id_F \otimes c)
=c\circ g
=\tilde{c}\circ y\circ g
=\tilde{c}\circ h\circ(\id_F\otimes y)~.
\ee
\end{proof}

\begin{Lem}\label{lem:CF-ker}
$\Cc_F$ has kernels.
\end{Lem}

\begin{proof}
We are given $U_f,V_g\in\Cc_F$ and a morphism $x:U_f\rightarrow V_g$. Since $\Cc$ has kernels, there exists an object $K \in \Cc$ and a morphism $\ker : K \rightarrow U$ such that $\ker$ is a kernel of $x$ in $\Cc$. We now wish to construct a morphism $k : F \otimes K \rightarrow K$ such that $\ker : K_h \rightarrow U_f$ is a morphism in $\Cc_F$. Consider the following diagram:
\be
\begin{xy}
(0,37)*+{F\otimes K}="fk"; (30,37)*+{F\otimes U}="fu"; (60,37)*+{F\otimes V}="fv";%
(0,20)*+{K}="k"; (30,20)*+{U}="u"; (60,20)*+{V}="v";%
{\ar@{-->} "fk";"k"};?*!/_1em/{\exists !h};{\ar "fu";"u"};?*!/_.6em/{f};{\ar "fv";"v"};?*!/_.6em/{g};%
{\ar "fk";"fu"}?*!/_.6em/{\id_F\otimes \ker};{\ar "fu";"fv"}?*!/_.6em/{\id_F\otimes x};
{\ar "k";"u"}?*!/_.6em/{\ker};{\ar "u";"v"}?*!/_.6em/{x};
\end{xy}
\ee
Note that $x\circ f\circ(\id_F\otimes \ker)=g\circ(\id_F\otimes(x\circ\ker))=0$. By the universal property of kernels in $\Cc$, there exists a unique morphism $h : F \otimes K \rightarrow K$ which makes the above diagram commute. Thus, $\ker : K_h \rightarrow U_f$ is a morphism in $\Cc_F$. Since $\ker$ is a kernel of $x$ in $\Cc$, by Lemma \ref{lem:CCF-ker-cok}\,(i) $\ker$ is also a kernel of $x$ in $\Cc_F$.
\end{proof}

\begin{Lem}\label{lem:CF-cok}
$\Cc_F$ has cokernels.
\end{Lem}

\begin{proof}
The proof is similar to that for the existence of kernels, with the difference that for the existence of cokernels we need the tensor product of $\Cc$ to be right-exact. We are given a morphism $x:U_f\rightarrow V_g$. The morphism $x$ has a cokernel $\cok : V \rightarrow C$ in $\Cc$. Consider the following diagram:
\be
\begin{xy}
(0,37)*+{F\otimes U}="fu"; (30,37)*+{F\otimes V}="fv"; (60,37)*+{F\otimes C}="fc";%
(0,20)*+{U}="u"; (30,20)*+{V}="v"; (60,20)*+{C}="c";%
{\ar "fu";"u"};?*!/_.6em/{f};{\ar "fv";"v"};?*!/_.6em/{g};{\ar@{-->} "fc";"c"};?*!/_1em/{\exists !c};%
{\ar "fu";"fv"}?*!/_.6em/{\id_F\otimes x};{\ar "fv";"fc"}?*!/_.6em/{\id_F\otimes \cok};
{\ar "u";"v"}?*!/_.6em/{x};{\ar "v";"c"}?*!/_.6em/{\cok};
\end{xy}
\ee
Since $\otimes$ is right-exact, $\id_F\otimes \cok$ is a cokernel of $\id_F\otimes x$. Note that $\cok\circ g\circ(\id_F\otimes x)=\cok\circ x\circ f=0$. By the universal property of cokernels in $\Cc$, there exists a unique morphism $c : F \otimes C \rightarrow C$ which makes the above diagram commute. Thus, $\cok : V_g \rightarrow C_c$ is a morphism in $\Cc_F$. Since $\cok$ is a cokernel of $x$ in $\Cc$, by Lemma \ref{lem:CCF-ker-cok}\,(ii) it is also a cokernel of $x$ in $\Cc_F$.
\end{proof}

The proof of Lemma \ref{lem:CF-ker} shows that there exists a kernel for $x : U_f \rightarrow V_g$ of the form $\ker : K_h \rightarrow U_f$, with $\ker$ a kernel of $x$ in $\Cc$. The proof of Lemma \ref{lem:CF-cok} implies a similar statement for cokernels. Since kernels and cokernels are unique up to unique isomorphism, we get as a corollary the converse statement to Lemma \ref{lem:CCF-ker-cok}.

\begin{Cor}\label{lem:CCF-ker-cok-conv}
Let $x : U_f \rightarrow V_g$ and $y : V_g \rightarrow W_h$ be morphisms in $\Cc_F$.\\
(i) If $x : U_f \rightarrow V_g$ is a kernel of $y$ in $\Cc_F$, then $x : U \rightarrow V$ is a kernel of $y$ in $\Cc$.\\
(ii) If $y : V_g \rightarrow W_h$ is a cokernel of $x$ in $\Cc_F$, then $y : V \rightarrow W$ is a cokernel of $x$ in $\Cc$.
\end{Cor}

We have now gathered the ingredients to prove Lemma \ref{lem:CF-exact}.

\medskip
\noindent{\em Proof of Lemma \ref{lem:CF-exact}.}
By Lemmas \ref{lem:CF-ker} and \ref{lem:CF-cok}, $\Cc_F$ has kernels and cokernels. Let $\chi : K_h \rightarrow V_g$ be a kernel of $b : V_g \rightarrow W_h$ and let $\gamma: V_g \rightarrow C_c$ be a cokernel of $a : U_f \rightarrow V_g$. By Corollary \ref{lem:CCF-ker-cok-conv}, also in $\Cc$ we have that
$\chi$ is a kernel of $b : V \rightarrow W$ and $\gamma$ is a cokernel of $a : U \rightarrow V$.

Suppose $U_f \xrightarrow{a} V_g \xrightarrow{b} W_h$ is exact at $V_g$ in $\Cc_F$, i.e.\ $\chi$ is also a kernel for $\gamma$ in $\Cc_F$. By Corollary \ref{lem:CCF-ker-cok-conv}, $\chi$ is a kernel for $\gamma$ in $\Cc$ and so $U \xrightarrow{a} V \xrightarrow{b} W$ is exact at $V$ in $\Cc$.
Conversely, if $\chi$ is a kernel for $\gamma$ in $\Cc$, then by Lemma \ref{lem:CCF-ker-cok} $\chi$ is also a kernel for $\gamma$ in $\Cc_F$. Thus $U_f \xrightarrow{a} V_g \xrightarrow{b} W_h$ is exact at $V_g$ in $\Cc_F$.
\hfill $\square$

\begin{Cor}\label{cor:monic-epi}
(to Lemma \ref{lem:CF-exact})
Let $x : U_f \rightarrow V_g$ be a morphism in $\Cc_F$. Then $x$ is monic in $\Cc_F$ iff it is monic in $\Cc$, and $x$ is epi in $\Cc_F$ iff it is epi in $\Cc$.
\end{Cor}

\begin{Lem}\label{lem:CF-bip}
$\Cc_F$ has binary biproducts.
\end{Lem}

\begin{proof}
Let $U_f, V_g \in \Cc_F$ be given.
Since $\Cc$ has binary biproducts, for $U,V\in\Cc$, there exists a $W\in\Cc$ and morphisms
\be
\begin{xy}
(0,20)*+{U}="u"; (20,20)*+{W}="w"; (40,20)*+{V}="v";%
{\ar@<1.ex>"u";"w"}?*!/_.6em/{e_U};{\ar@<-1.ex>"v";"w"}?*!/^2mm/{e_V};%
{\ar@<1.ex>"w";"u"}?*!/_.6em/{r_U};{\ar@<-1.ex>"w";"v"}?*!/^2mm/{r_V};%
\end{xy}
\labl{eq:ab-binprod-pf1}
where $e_A$ is the embedding map and $r_A$ is the restriction map, such that
\be
    r_U\circ e_U=\id_U, \ \ r_V\circ e_V=\id_V, \ \ e_U\circ r_U+e_V\circ r_V=\id_W.
\ee
This implies $r_U\circ e_V=0$ and $r_V\circ e_U=0$. Define a morphism $h : F \otimes W \rightarrow W$ as
\be
  h=e_U\circ f\circ(\id_F\otimes r_U)+e_V\circ g\circ(\id_F\otimes r_V) ~.
\ee
We claim that \eqref{eq:ab-binprod-pf1} with $U$, $W$ and $V$ replaced by $U_f$, $W_h$ and $V_g$, respectively, defines a binary biproduct in $\Cc_F$.
To show these we need to check that the relevant four squares in
\be
\begin{xy}
(0,20)*+{F\otimes U}="fu"; (40,20)*+{F\otimes W}="fw"; (80,20)*+{F\otimes V}="fv";%
{\ar@<1.ex>"fu";"fw"}?*!/_.6em/{\id_F\otimes e_U};{\ar@<-1.ex>"fv";"fw"}?*!/^2mm/{\id_F\otimes e_V};%
{\ar@<1.ex>"fw";"fu"}?*!/_.6em/{\id_F\otimes r_U};{\ar@<-1.ex>"fw";"fv"}?*!/^2mm/{\id_F\otimes r_V};%
(0,0)*+{U}="u"; (40,0)*+{W}="w"; (80,0)*+{V}="v";%
{\ar@<1.ex>"u";"w"}?*!/_.6em/{e_U};{\ar@<-1.ex>"v";"w"}?*!/^2mm/{e_V};%
{\ar@<1.ex>"w";"u"}?*!/_.6em/{r_U};{\ar@<-1.ex>"w";"v"}?*!/^2mm/{r_V};%
{\ar "fu";"u"}?*!/_.6em/{f};{\ar "fw";"w"}?*!/_.6em/{h};{\ar "fv";"v"}?*!/_.6em/{g};%
\end{xy}
\ee
commute. For the first square one has
\bea
h\circ(\id_F\otimes e_U)
=e_U\circ f\circ(\id_F\otimes (\underbrace{r_U\circ e_U}_{=\id_U}))+e_V\circ g\circ(\id_F\otimes (\underbrace{r_V\circ e_U}_{=0}))
=e_U\circ f ~,
\\[-1.5em]
\eear\ee
and for the second one
\be
r_U\circ h =r_U\circ e_U\circ f\circ(\id_F\otimes r_U)+r_U\circ e_V\circ g\circ(\id_F\otimes r_V)=f\circ(\id_F\otimes r_U)~.
\ee
In a similar fashion one checks that also $h\circ(\id_F\otimes e_V)=e_V\circ g$ and $r_V\circ h=g\circ(\id_F\otimes r_V)$.
\end{proof}

\begin{Lem}\label{lem:CF-mon-ep}
In $\Cc_F$ every monomorphism is a kernel and every epimorphism is a cokernel.
\end{Lem}

\begin{proof}
First we show that every monomorphism is a kernel. We need to show that if $x:U_f \rightarrow V_g$ is mono in $\Cc_F$, there exists a $W_h$ and $y:V_g \rightarrow W_h$ such that $x=\ker y$. Since $\Cc_F$ has cokernels we can choose $W_h=C_c$ and $y=\cok x$. Since by Corollary \ref{cor:monic-epi} $x$ is monic also in $\Cc$, we have $x=\ker(\cok x)$ in $\Cc$. Finally, by Lemma \ref{lem:CCF-ker-cok} we get that $x=\ker(\cok x)$ also in $\Cc_F$. The proof that every epimorphism is a cokernel goes along the same lines.
\end{proof}

\medskip
\noindent{\em Proof of Theorem \ref{thm:ab}.}
Since $\Cc$ is an Ab-category, so is $\Cc_F$. As zero object in $\Cc_F$ we take $({\bf 0},0)$, where ${\bf 0}$ is the zero object of $\Cc$ and $0:F\otimes 0\rightarrow 0$ is the zero morphism. Furthermore, $\Cc_F$ has binary biproducts (Lemma \ref{lem:CF-bip}), has kernels and cokernels (Lemmas \ref{lem:CF-ker} and \ref{lem:CF-cok}) and in $\Cc_F$ every monomorphism is a kernel and every epimorphism is a cokernel (Lemma \ref{lem:CF-mon-ep}). Thus $\Cc_F$ is abelian.
\hfill $\square$

\subsection{Finite semi-simple monoidal categories}\label{app:catV}

Let $\Bbbk$ be a field. In this section we take $\Cc$ to be a $\Bbbk$-linear abelian semi-simple finite braided monoidal category, such that $\one$ is simple, and $\End(U) = \Bbbk \id_U$ for all simple objects $U$. We also assume that $\Cc$ has right duals and that
$$
  \Cc \text{ is strict.}
$$
Note that if we would add to this the data/conditions that $\Cc$ has compatible left-duals and a twist (so that $\Cc$ is ribbon), we would arrive at the definition of a premodular category \cite{brug2}. Here we will content ourselves with right duals alone.

For explicit calculations in $\Cc_F$ it is useful to have a realisation of $\Cc$ in terms of vector spaces. One way to obtain such a realisation is as follows.
Pick a set of representatives $\{ U_i | i \in \Ic \}$ of the isomorphism classes of simple objects in $\Cc$ such that $U_0 = \one$. For each label $a \in \Ic$ define a label $\bar a$ via $U_{\bar a} \cong U_a^\vee$. Define the fusion rule coefficients $N_{ij}^{~k}$ as
\be
  N_{ij}^{~k} = \dim_\Bbbk \Hom(U_i \otimes U_j, U_k) ~.
\ee
We restrict ourselves to the situation that
\be
  N_{ij}^{~k} \in \{0,1\} ~.
\ee
This is satisfied in the Lee-Yang model studied below, but also for other models such as the rational free boson or the $\widehat{su}(2)_k$-WZW model. Whenever $N_{ij}^{~k}=1$ we pick basis vectors
\be
  \lambda_{(ij)k} \in \Hom(U_i \otimes U_j, U_k)
  \quad \text{such that} \quad
  \lambda_{(0i)i} = \lambda_{(i0)i} = \id_{U_i} ~.
\ee
The fusing matrices $\Fs^{(ijk)l}_{pq} \in \Bbbk$ are defined to implement the change of basis between two bases of $\Hom(U_i \otimes U_j \otimes U_k,U_l)$ as follows,
\be
  \lambda_{(ip)l} \circ (\id_{U_i} \otimes \lambda_{(jk)p})
  = \sum_{q \in \Ic} \Fs^{(ijk)l}_{pq} \cdot \lambda_{(qk)l} \circ (\lambda_{(ij)q} \otimes \id_{U_k}) ~.
\ee
The fusing matrices obey the pentagon relation. See e.g.\ \cite[Sect.\,2.2]{tft1} for a graphical representation and more details. The inverse matrices are denoted by $\Gs^{(ijk)l}_{pq}$,
\be
  \sum_{r \in \Ic} \Fs^{(ijk)l}_{pr} \Gs^{(ijk)l}_{rq} = \delta_{p,q} ~.
\ee
The braiding $c_{U,V}$ gives rise to the braid matrices $\Rs^{(ij)k} \in \Bbbk$,
\be
  \lambda_{(ji)k} \circ c_{U_i,U_j} = \Rs^{(ij)k} \, \lambda_{(ij)k} ~.
\ee

With these ingredients, we define a $\Bbbk$-linear braided monoidal category $\Vc \equiv \Vc[\Bbbk,\Ic,0\in\Ic,N,\Fs,\Rs]$. This definition will occupy the rest of this section. The objects of $\Vc$ are lists of finite-dimensional $\Bbbk$-vector spaces indexed by $\Ic$, $A = (A_i, i \in \Ic)$, and the morphisms $f : A \rightarrow B$ are lists of linear maps $f = (f_i , i \in \Ic)$ with $f_i : A_i \rightarrow B_i$.

There is an obvious functor $H : \Cc \rightarrow \Vc$ which acts on objects as  $H(V) = ( \Hom(U_i,V) , i \in \Ic)$. For a morphism $f : V \rightarrow W$ we set $H(f) = (H(f)_i , i \in \Ic)$, where $H(f)_i : \Hom(U_i,V) \rightarrow \Hom(U_i,W)$ is given by $\alpha \mapsto f \circ \alpha$. Since $H$ is fully faithful and surjective we have:

\begin{Lem}
The functor $H : \Cc \rightarrow \Vc$ is an equivalence of $\Bbbk$-linear categories.
\end{Lem}

We can now use $H$ to transport the tensor product, braiding and duality from $\Cc$ to $\Vc$. Let us start with the tensor product in $\Vc$, which we denote by $\circledast$. For an object $A \in \Vc$ we denote by $(A)_i$ (or just $A_i$) the $i$'th component of the list $A$. We set
\be
  (A \circledast B)_i = \bigoplus_{j \in \Ic}  \bigoplus_{k \in \Ic, N_{jk}^{~i}=1} A_j \otimes_\Bbbk B_k ~.
\ee
The direct summand $A_j \otimes_\Bbbk B_k$ can appear in several components $(A \circledast B)_i$. To index one specific direct summand, we introduce the notation $(A \circledast B)_{i(jk)}$ to mean
\be
  (A \circledast B)_{i(jk)} = A_j \otimes_\Bbbk B_k \subset  (A \circledast B)_i ~.
\labl{eq:V-tens-subnot}
This notation can be iterated. For example $( A \circledast ( B \circledast C) )_{i(jk(lm))}$ stands for the direct summand (we do not write out the associator and unit isomorphisms in the category of $\Bbbk$-vector spaces)
\be
  A_j \otimes_\Bbbk B_l \otimes_\Bbbk C_m
  \subset A_j \otimes_\Bbbk (B \circledast C)_k \subset ( A \circledast ( B \circledast C) )_i ~.
\ee
while
$( (A \circledast B) \circledast C )_{i(j(kl)m)}$ stands for the direct summand
\be
  A_k \otimes_\Bbbk B_l \otimes_\Bbbk C_m
  \subset (A \circledast B)_j \otimes_\Bbbk C_m \subset ( (A \circledast  B) \circledast C )_i ~.
\ee
If $v \in A_j \otimes_\Bbbk B_k$, we denote by $(v)_{i(jk)}$ the element $v$ in the direct summand $(A \circledast B)_{i(jk)} \subset (A \circledast B)_i$, etc.

On morphisms $f : A \rightarrow X$ and $g : B \rightarrow Y$ the tensor product is defined to have components $( f \circledast g)_i : (A \circledast B)_i \rightarrow (X \circledast Y)_i$, where, for $a \in A_j$ and $b \in B_k$,
\be
  \big( f \circledast g\big)_i\big( (a \otimes_\Bbbk b)_{i(jk)}\big) = \big( \, f_j(a) \otimes_\Bbbk g_k(b) \, \big)_{i(jk)}
  ~ \in X_j \otimes_\Bbbk Y_k \subset (X \circledast Y)_i ~.
\ee
The tensor unit $\one \in \Vc$ has components $\one_0 = \Bbbk$ and $\one_i = 0$ for $i \neq 0$. The unit isomorphisms of $\Vc$ are identities, but we find it useful to write them out to keep track of the indices of the direct summands,
\be
\begin{array}{rll}\displaystyle
  (\lambda_A)_i ~:~~ (\one \circledast A)_i ~~ \etb\longrightarrow\etb A_i
  \\
  (1 \otimes_\Bbbk a)_{i(0i)} \etb \longmapsto\etb (a)_i
\eear
\qquad \text{and} \qquad
\begin{array}{rll}\displaystyle
  (\rho_A)_i ~:~~ (A \circledast \one)_i ~~ \etb\longrightarrow\etb A_i
  \\
  (a \otimes_\Bbbk 1)_{i(i0)} \etb \longmapsto\etb (a)_i
\eear
\quad .
\ee
Finally, the associator has components $(\alpha_{A,B,C})_i : ( A \circledast ( B \circledast C) )_i \rightarrow ( (A \circledast B ) \circledast C )_i$, where, for $v \in A_j \otimes_\Bbbk B_k \otimes_\Bbbk C_l$,
\be
  (\alpha_{A,B,C})_i\big( \, (v)_{i(jq(kl))} \, \big)
  = \sum_{p \in \Ic} \big( \, \Gs^{(jkl)i}_{pq} \, v\,  \big)_{i(p(jk)l)} ~.
\labl{eq:V-assoc}
Its inverse is $(\alpha_{A,B,C})_i^{-1} : ( ( A \circledast B)  \circledast C )_i \rightarrow ( A \circledast ( B \circledast C ) )_i$,
\be
  (\alpha_{A,B,C}^{-1})_i\big( \, (v)_{i(q(jk)l)} \, \big)
  = \sum_{p \in \Ic} \big( \, \Fs^{(jkl)i}_{pq} \, v\,  \big)_{i(jp(kl))} ~.
\labl{eq:V-inv-assoc}
We can now turn $H$ into a monoidal functor. To this end we need to specify natural isomorphisms
$H^2_{U,V} : H(U) \circledast H(V) \rightarrow H(U \otimes V)$ and an isomorphism $H^0 : \one_\Vc \rightarrow H(\one_\Cc)$.  To describe $H^2_{U,V}$ we need the basis dual to $\lambda_{(ij)k}$, that is, elements $y_{(ij)k} \in \Hom(U_k,U_i \otimes U_j)$ such that $\lambda_{(ij)k} \circ y_{(ij)k} = \id_{U_k}$. Note that $(H(U) \circledast H(V))_{i(jk)} = \Hom(U_j,U) \otimes_\Bbbk \Hom(U_k,V)$ and $H(U \otimes V)_i = \Hom(U_i,U\otimes V)$. We set, for $u \in \Hom(U_j,U)$ and $v \in \Hom(U_k,V)$,
\be
  (H^2_{U,V})_i( (u \otimes_\Bbbk v)_{i(jk)} ) = ( (u \otimes v) \circ y_{(jk)i} )_i ~.
\ee
Finally, $(H^0)_i =0$ for $i\neq 0$ and $(H^0)_0(1) = \id_{U_0} \in \Hom(U_0,U_0)$.

\begin{Thm}
$(H,H^2,H^0) : \Cc \rightarrow \Vc$ is a monoidal functor.
\end{Thm}

\begin{proof}
We have to check that for all $U,V,W \in \Cc$ the following equalities of morphisms
$H(U) \circledast( H(V) \circledast H(W)) \rightarrow H(U \otimes V \otimes W)$,
$\one_\Vc \circledast H(U) \rightarrow H(U)$
and
$H(U) \circledast \one_\Vc \rightarrow H(U)$,
respectively, hold,
\bea
  H^2_{U \otimes V,W} \circ ( H^2_{U,V} \circledast \id_{H(W)} ) \circ \alpha_{H(U),H(V),H(W)}
  = H^2_{U, V\otimes W} \circ (\id_{H(U)} \circledast H^2_{V,W}) ~,
\enl
  \lambda_{H(U)} = H^2_{H(\one),H(U)} \circ (H^0 \circledast \id_{H(U)}) ~~,~~~
  \rho_{H(U)} = H^2_{H(U),H(\one)} \circ (\id_{H(U)} \circledast H^0) ~.
\eear\labl{eq:H-mon-fun-cond}
(Recall that $\Cc$ is strict.) The identities involving $\lambda$ and $\rho$ are most easy to check. For example, the $i$th component of two sides of the identity for $\lambda$ are, for $u \in \Hom(U_i,U)$,
\bea
  (\lambda_{H(U)})_i( (1 \otimes_\Bbbk u)_{i(0i)} ) = (u)_i \qquad \text{and}
  \enl
  (H^2_{H(\one),H(U)})_i \circ (H^0 \circledast \id_{H(U)})_i(  (1 \otimes_\Bbbk u)_{i(0i)} )
  =   (H^2_{H(\one),H(U)})_i( (\id_{U_0} \otimes_\Bbbk u)_{i(0i)} )
  \\[.3em]\displaystyle
  =  ( (\id_{U_0} \otimes u) \circ y_{(0i)i} )_i = (u)_i ~.
\eear\ee
To check the first condition in \eqref{eq:H-mon-fun-cond} we pick elements
$u \in \Hom(U_j,U)$, $v \in \Hom(U_k,V)$, $w \in \Hom(U_l,W)$ and evaluate both sides on the element
$(u \otimes_\Bbbk v  \otimes_\Bbbk w)_{i(jq(kl))}$. For the left hand side this gives
\bea
  \big(H^2_{U \otimes V,W} \circ ( H^2_{U,V} \circledast \id_{H(W)} ) \circ \alpha_{H(U),H(V),H(W)}\big)_i\big((u \otimes_\Bbbk v  \otimes_\Bbbk w)_{i(jq(kl))}\big)
  \enl
  = \sum_{p \in \Ic}
  \big(H^2_{U \otimes V,W} \circ ( H^2_{U,V} \circledast \id_{H(W)} ) \big)_i\big( (\Gs^{(jkl)i}_{pq} \cdot  u \otimes_\Bbbk v  \otimes_\Bbbk w)_{i(p(jk)l)}\big)
  \enl
  = \sum_{p \in \Ic}
  (H^2_{U \otimes V,W})_i\big(  ( \Gs^{(jkl)i}_{pq} \cdot  ((u \otimes v) \circ y_{(jk)p}) \otimes_\Bbbk w)_{i(pl)}\big)
  \enl
  = \big( \sum_{p \in \Ic}  \Gs^{(jkl)i}_{pq} \cdot  (((u \otimes v) \circ y_{(jk)p}) \otimes w) \circ y_{(pl)i}  \big)_i
  = \big( (u \otimes v \otimes w) \circ (\id_{U_j} \otimes y_{(kl)q}) \circ y_{(jq)i}  \big)_i ~.
\eear\ee
For the right hand side we find
\bea
\big( H^2_{U, V\otimes W} \circ (\id_{H(U)} \circledast H^2_{V,W}) \big)_i\big((u \otimes_\Bbbk v  \otimes_\Bbbk w)_{i(jq(kl))}\big)
\enl
= \big( H^2_{U, V\otimes W} \big)_i\big( (u \otimes_\Bbbk [ (v \otimes w) \circ y_{(kl)q}])_{i(jq)} \big)
\enl
= \big( (u \otimes [ (v \otimes w) \circ y_{(kl)q}]) \circ y_{(jq)i} \big)_i
= \big( (u \otimes v \otimes w) \circ (\id_{U_j} \otimes y_{(kl)q}) \circ y_{(jq)i}  \big)_i ~.
\eear\ee
Thus $H$ is indeed a monoidal functor.
\end{proof}

We define a braiding $c_{A,B} : A \circledast B \rightarrow B \circledast A$ on $\Vc$ by setting, for $a \in A_j$ and $b \in B_k$,
\be
  (c_{A,B})_i( (a \otimes b)_{i(jk)}) = ( \Rs^{(jk)i} b \otimes a )_{i(kj)} ~.
\labl{eq:V-braid}
One verifies that $H(c_{U,V}) \circ H^2_{U,V} = H^2_{V,U} \circ c_{H(U),H(V)}$ so that $H$ provides a braided monoidal equivalence between $\Cc$ and $\Vc$.

It remains to define the right duality on $\Vc$. The components of the dual of an object are given by dual vector spaces, $(A^\vee)_k = A^*_{\bar k}$. We identify $\Bbbk^* = \Bbbk$ so that $\one^\vee = \one$. The duality morphisms $b_A : \one \rightarrow A \circledast A^\vee$ and $d_A : A^\vee \circledast A \rightarrow \one$ have components $(b_A)_i = 0 = (d_A)_i$ for $i \neq 0$. To describe the 0-component, we fix a basis $\{ a_{i,\alpha} \}$ of each $A_i$, and denote by $\{ a_{i,\alpha}^* \}$ the dual basis of $A_i^*$. Then
\be
\begin{array}{rll}\displaystyle
  (b_A)_0 : (\one)_0 \etb\longrightarrow\etb (A \circledast A^\vee)_0
  \\
  (1)_0 \etb \longmapsto\etb
  \sum_{k \in \Ic}\Big( \sum_{\alpha} a_{k,\alpha} \otimes_\Bbbk a_{k,\alpha}^* \Big)_{0(k \bar k)}
\eear
~,~
\begin{array}{rll}\displaystyle
  (d_A)_0 : (A^\vee \circledast A)_0 \etb\longrightarrow\etb (\one)_0
  \\
  (\varphi \otimes_\Bbbk a)_{0(\bar kk)} \etb \longmapsto\etb
  \frac{\varphi(a)}{\Fs^{(k\bar kk)k}_{00}}
\eear
~.
\ee
As an exercise in the use of the nested index notation we demonstrate the second identity in \eqref{eq:r-dual}. Let $a_{i,\alpha}$, $a_{i,\alpha}^*$ be as above. Then, for $\varphi \in A^*_{\bar k}$,
\be\begin{array}{rl}
\big(\, \rho^{-1}_{A^\vee} \,\big)_k\big(\, (\varphi)_{k} \,\big) \etb= (\varphi \otimes_\Bbbk 1)_{k(k0)} = \star_1
\enl
\big(\, \id_{A^\vee} \circledast b_A \,\big)_k(\star_1) \etb=
  \sum_{l \in \Ic} \sum_\alpha \big(\, (\varphi)_k \otimes_\Bbbk ( a_{k,\alpha} \otimes_\Bbbk a_{k,\alpha}^* )_{0(l \bar l)} \,\big)_{k(k0)}
  \enl\etb=
  \sum_{l,\alpha} \big(\, \varphi \otimes_\Bbbk a_{k,\alpha} \otimes_\Bbbk a_{k,\alpha}^* \,\big)_{k(k0(l \bar l))} = \star_2
\enl
\big(\, \alpha_{A^\vee,A,A^\vee} \,\big)_k(\star_2) \etb=
\sum_{p \in \Ic} \sum_{l,\alpha}  \big(\, \Gs^{(kl\bar l)k}_{p0} \cdot \varphi \otimes_\Bbbk a_{k,\alpha} \otimes_\Bbbk a_{k,\alpha}^* \,\big)_{k(p(kl)\bar l)}
= \star_3
\enl
\big(\, d_A \circledast id_{A^\vee} \,\big)_k(\star_3) \etb=
\sum_{p,l,\alpha}  \big(\, \Gs^{(kl\bar l)k}_{p0} \cdot (d_A)_p( (\varphi \otimes_\Bbbk a_{k,\alpha})_{p(kl)}) \otimes_\Bbbk (a_{k,\alpha}^*)_{\bar l} \,\big)_{k(p\bar l)}
\enl\etb\overset{\text{(a)}}=
\sum_{\alpha}  \big(\, \Gs^{(k\bar kk)k}_{00} (\Fs^{(\bar kk\bar k)\bar k}_{00})^{-1} \cdot \varphi(a_{k,\alpha}) \otimes_\Bbbk a_{k,\alpha}^* \,\big)_{k(0k)}
\enl\etb\overset{\text{(b)}}=
\big(\, 1 \otimes_\Bbbk \varphi \,\big)_{k(0k)}
= \star_4
\enl
\big(\, \lambda_{A^{\vee}} \,\big)_k(\star_4) \etb=(\varphi)_{k} ~.
\eear\ee
In step (a) we used that $(d_A)_p$ is non-zero only for $p=0$, and that in this case we are also forced to choose $l = \bar k$ (otherwise the direct summand $(\cdots)_{0(kl)}$ is empty). In step (b) the equality
\be
  \Fs^{(\bar kk\bar k)\bar k}_{00} = \Gs^{(k\bar kk)k}_{00}
\labl{eq:F00=G00}
is used. This equality can be derived by using either $\Fs$ or $\Gs$ to simplify $(\lambda_{(\bar kk)0} \otimes \lambda_{(\bar kk)0}) \circ (\id_{U_{\bar k}} \otimes y_{(k\bar k)0} \otimes \id_{U_k})$ to $\lambda_{(\bar kk)0}$ (which also shows that both are non-zero).

\begin{Rem}
(i) The above construction is a straightforward generalisation of the way one defines a (braided) monoidal category starting from a (abelian) group and a (abelian) three-cocycle, see \cite[Sect.\,2]{tft3} and references therein.
\\
(ii) The construction is different from what one would do in Tannaka-Krein reconstruction for monoidal categories \cite{Hayashi:1999}. There one constructs a fibre-functor from $\Cc$ to a category of $R$-$R$-bimodules for a certain ring $R$ (isomorphic to $\Bbbk^{\oplus |\Ic|}$). However, this fibre-functor is typically neither an equivalence nor full.
\end{Rem}

Let $f : F \circledast A \rightarrow A$ and $g : F \circledast B \rightarrow B$ be morphisms in $\Vc$.
We can now substitute the explicit structure morphisms \eqref{eq:V-assoc}, \eqref{eq:V-inv-assoc}, \eqref{eq:V-braid} into the definition of $T(f,g)$ in Section \ref{sec:CF-mon}. After a short calculation one finds, for $u \in F_j$, $a \in A_l$ and $b \in B_m$,
\bea
  T(f,g)_i\big(\, (u \otimes_\Bbbk a \otimes_\Bbbk b)_{i(jk(lm))} \,\big)
  \enl
  \qquad
  = \sum_{x,y \in \Ic} \Big(
  \delta_{y,m} \Gs^{(jlm)i}_{xk}
  (f)_x\big( (u \otimes_\Bbbk a)_{x(jl)} \big) \otimes_\Bbbk (b)_y
  \enl
  \hspace{7em}
  +~
  \delta_{x,l} \frac{\Rs^{(jk)i}}{\Rs^{(jm)y}} \Fs^{(lmj)i}_{yk} (a)_x \otimes_\Bbbk (g)_y\big( (u\otimes_\Bbbk b)_{y(jm)} \big) \Big)_{\! i(xy)} ~.
\eear\labl{eq:T-in-V}
When verifying this one needs to use the following two equivalent expressions for the $\mathsf{B}$-matrix (see e.g.\ \cite[Eqn.\,(5.46)]{tft4}), one of which is \cite[Eqn.\,(5.47)]{tft4} and the other one appears in the calculation of $T(0,g)_i\big(\, (u \otimes_\Bbbk a \otimes_\Bbbk b)_{i(jk(lm))} \,\big)$,
\be
  \sum_p \Fs^{(ljm)i}_{yp} \, \Rs^{(jl)p} \, \Gs^{(jlm)i}_{pk} ~=~ \mathsf{B}^{(jlm)i}_{yk} \,=~ \frac{\Rs^{(jk)i}}{\Rs^{(jm)y}} \, \Fs^{(lmj)i}_{yk} ~.
\ee
For $c(f)$ the calculation is slightly longer, and one finds, for $u \in F_j$ and $\varphi \in A^*_{\bar k}$, and using \eqref{eq:F00=G00} at an intermediate step,
\bea
  c(f)_i\big(\, (u \otimes_\Bbbk \varphi)_{i(jk)}\, \big)
  \enl
  \qquad = - \frac{\Fs^{(\bar\imath i\bar\imath) \bar\imath}_{00}}{\Fs^{(\bar kk\bar k) \bar k}_{00}} \,
  \Rs^{(jk)i} \, \Fs^{(kj\bar\imath)0}_{\bar k i} \sum_\alpha \varphi\big(
  (f)_{\bar k}( (u \otimes a_{\bar\imath,\alpha})_{\bar k(j \bar \imath)} ) \big) \cdot a_{\bar\imath,\alpha}^*
  ~~ \in ~ (A^\vee)_i = A^*_{\bar\imath} ~.
\eear\labl{eq:c-in-V}

\subsection[$T(f,g)$ and $c(f)$ for the Lee-Yang model]{$\boldsymbol{T(f,g)}$ and $\boldsymbol{c(f)}$ for the Lee-Yang model} \label{app:T-c-LY}

The Lee-Yang model is the minimal model $M(2,5)$. The fusing matrices of minimal models are known from \cite{Dotsenko:1984ad,Furlan:1989ra}. We use the conventions of \cite[App.\,A.3]{Runkel:2007wd}. The index set is $\Ic = \{1,\phi\}$ and the unit element is $1 \in \Ic$. The non-zero entries in the braiding matrix are, for $x \in \{ 1, \phi \}$
\be
  \Rs^{(\one x)x} =   \Rs^{(x \one)x} = 1 ~~,~~~
  \Rs^{(\phi \phi)1} = \zeta^2 ~~,~~~
  \Rs^{(\phi \phi)\phi} = \zeta ~~,~~~
  \text{where} ~
  \zeta = e^{- \pi i /5} ~.
\ee
The nonzero entries in the fusing matrices are, for $x,y,z \in \{ 1, \phi \}$
\bea
  \Fs^{(1xy)z}_{zx} =
  \Fs^{(x1y)z}_{yx} =
  \Fs^{(xy1)z}_{yz} =
  \Fs^{(xyz)1}_{xz} = 1 ~,
  \enl
  \Fs^{(\phi\phi\phi)\phi}_{11} = \frac{1}{d} ~,~~
  \Fs^{(\phi\phi\phi)\phi}_{1\phi} = w ~,~~
  \Fs^{(\phi\phi\phi)\phi}_{\phi1} = \frac{1}{wd}~,~~
  \Fs^{(\phi\phi\phi)\phi}_{\phi\phi} = \frac{-1}{d}
  ~~\text{where}~  d = \frac{1-\sqrt{5}}2 ~.
\eear\ee
Here $d$ is the quantum dimension of $\phi$. The constant $w \in \Cb^\times$ depends on the choice of normalisation of the basis vectors $\lambda_{(\phi\phi)1}$ and $\lambda_{(\phi\phi)\phi}$. Different choices of $w$ yield equivalent braided monoidal categories. There is a preferred choice related to the normalisation of the vertex operators, for which
\be
  w = \frac{\Ga{\tfrac15}\Ga{\tfrac65}}{\Ga{\tfrac35}\Ga{\tfrac45}} = 2.431... ~,
\ee
but one may as well set $w$ to $1$. The inverse matrix of $\Fs$ is simply
\be
  \Gs^{(ijk)l}_{pq} = \Fs^{(kji)l}_{pq} ~.
\ee
Let us indicate how to obtain the explicit formulas quoted in Section \ref{sec:LY-ex}. First of all, in terms of the notation \eqref{eq:V-tens-subnot} for the direct summands of $A \circledast B$, the individual components in \eqref{eq:LY-AB-tens} are, in the same order,
\bea
  A \circledast B =
  \big((A \circledast B)_1,(A \circledast B)_\phi\big)
  \\[.3em]\displaystyle
  = \big( ~
  (A \circledast B)_{1(11)} \oplus
  (A \circledast B)_{1(\phi\phi)} ~,~
  (A \circledast B)_{\phi(1\phi)} \oplus
  (A \circledast B)_{\phi(\phi 1)} \oplus
  (A \circledast B)_{\phi(\phi\phi)} ~ \big) ~.
\eear\ee
Consider a morphism $f : \Phi \circledast A \rightarrow A$. In terms of three linear maps in \eqref{eq:LY-f-PhiA-A} the action of $f$ on the individual summands of $\Phi \circledast A$ is as 
follows. For $1 \in \Phi_\phi=\Cb$, $a \in A_1$ and $b \in A_\phi$,
\be\begin{array}{ll}\displaystyle
  & (f)_1\big(\,(1 \otimes_\Cb b)_{1(\phi\phi)}\,\big) = f_{1\phi}(b) ~,
\\[.3em]\displaystyle
  (f)_\phi\big(\,(1 \otimes_\Cb a)_{\phi(\phi1)}\,\big) = f_{\phi1}(a) ~,~&
  (f)_\phi\big(\,(1 \otimes_\Cb b)_{\phi(\phi\phi)}\,\big) = f_{\phi\phi}(b) ~.
\eear\ee
To obtain the expression \eqref{eq:LY-dual-c} for the dual of an object in $\Vc_\Phi$ we have to specialise \eqref{eq:c-in-V} to the Lee-Yang model. For example,
for $f : \Phi \circledast A \rightarrow A$ and $\varphi \in A_\phi^*$ one gets
\bea
  c(f)_1\big( \, (1 \otimes_\Cb \varphi)_{1(\phi\phi)}\,\big)
  = - \frac{\Fs^{(111)1}_{11}}{\Fs^{(\phi\phi\phi)\phi}_{11}} \,
  \Rs^{(\phi\phi)1} \, \Fs^{(\phi\phi1)1}_{\phi 1} \sum_\alpha \varphi\big(
  (f)_{\phi}( (u \otimes a_{1,\alpha})_{\phi(\phi 1)} ) \big) \cdot a_{1,\alpha}^*
\enl
  = - d \zeta^2 \sum_\alpha \varphi\big( f_{\phi 1}(a_{1,\alpha}) \big) \cdot a_{1,\alpha}^*
  = - d \zeta^2 f_{\phi 1}^*(\varphi) ~,
\eear\ee
which is the top right corner in \eqref{eq:LY-dual-c}. Expression \eqref{eq:LY-Tfg} for the tensor product of two morphisms in $\Vc_\Phi$ is obtained from \eqref{eq:T-in-V}. Denote by $T_{i(\phi k(lm))}^{i(xy)}$ the linear map $T(f,g)_i$ restricted to $(\Phi \circledast (A \circledast B))_{i(\phi k(lm))}$ and projected to the summand $(A \circledast B)_{i(xy)}$,
\be
  T_{i(\phi k(lm))}^{i(xy)} =
  \delta_{y,m} \Fs^{(ml\phi)i}_{xk} f_{x l} \otimes_\Cb \id_{B_y}
  + \delta_{x,l} \frac{\Rs^{(\phi k)i}}{\Rs^{(\phi m)y}} \Fs^{(lm\phi )i}_{yk} \id_{A_x} \otimes_\Cb g_{ym}
\ee
In terms of these, the elements of the matrix \eqref{eq:LY-Tfg} are
\be
  \begin{array}{cc}
    & \!\!\!\!
    \begin{array}{ccccc}
    A_1  B_1  ~~&~~
    A_\phi  B_\phi ~~&~~
    A_1  B_\phi  ~~&~~
    A_\phi  B_1  ~~&~~
    A_\phi  B_\phi
    \end{array} \\[.8em]
    \begin{array}{c}
    A_1  B_1
    \\[.8em]
    A_\phi  B_\phi
    \\[.8em]
    A_1  B_\phi
    \\[.8em]
    A_\phi  B_1
    \\[.8em]
    A_\phi  B_\phi
    \end{array}
    &
    \!\!\!\! \begin{pmatrix}
    0 & 0 & T_{1(\phi \phi(1\phi))}^{1(11)} & T_{1(\phi \phi(\phi1))}^{1(11)} & T_{1(\phi \phi(\phi\phi))}^{1(11)}
    \\[.8em]
    0 & 0 & T_{1(\phi \phi(1\phi))}^{1(\phi\phi)} & \underline{T_{1(\phi \phi(\phi1))}^{1(\phi\phi)}} & \underline{T_{1(\phi \phi(\phi\phi))}^{1(\phi\phi)}}
    \\[.8em]
    T_{\phi(\phi 1(11))}^{\phi(1\phi)} & T_{\phi(\phi 1(\phi\phi))}^{\phi(1\phi)} & T_{\phi(\phi \phi(1\phi))}^{\phi(1\phi)} & T_{\phi(\phi \phi(\phi 1))}^{\phi(1\phi)} & T_{\phi(\phi \phi(\phi\phi))}^{\phi(1\phi)}
    \\[.8em]
    T_{\phi(\phi 1(11))}^{\phi(\phi 1)} & T_{\phi(\phi 1(\phi\phi))}^{\phi(\phi 1)} & T_{\phi(\phi \phi(1\phi))}^{\phi(\phi 1)} & T_{\phi(\phi \phi(\phi 1))}^{\phi(\phi 1)} & T_{\phi(\phi \phi(\phi\phi))}^{\phi(\phi 1)}
    \\[.8em]
    T_{\phi(\phi 1(11))}^{\phi(\phi\phi)} & \underline{T_{\phi(\phi 1(\phi\phi))}^{\phi(\phi\phi)}} & T_{\phi(\phi \phi(1\phi))}^{\phi(\phi\phi)} & T_{\phi(\phi \phi(\phi 1))}^{\phi(\phi\phi)} & T_{\phi(\phi \phi(\phi\phi))}^{\phi(\phi\phi)}
    \end{pmatrix}
  \end{array}
\ee
For example, the underlined entries are
\bea
  T_{1(\phi \phi(\phi1))}^{1(\phi\phi)} = \zeta^2 \cdot \id_{A_\phi} \otimes_\Cb \, g_{\phi 1}
  ~,\\[.4em]\displaystyle
  T_{1(\phi \phi(\phi\phi))}^{1(\phi\phi)} =  f_{\phi \phi} \otimes_\Cb \, \id_{B_\phi} + \zeta \cdot \id_{A_\phi} \otimes_\Cb \, g_{\phi \phi}
  ~,\\[.4em]\displaystyle
  T_{\phi(\phi 1(\phi\phi))}^{\phi(\phi\phi)} = \tfrac{1}{wd} \cdot f_{\phi \phi} \otimes_\Cb \, \id_{B_\phi} + \tfrac{1}{\zeta wd} \cdot \id_{A_\phi} \otimes_\Cb \, g_{\phi \phi}
  ~,
\eear\ee
in agreement with \eqref{eq:LY-Tfg}.

\end{document}